\documentclass[prd,preprint,nofootinbib]{revtex4-1}

\usepackage{amssymb} 
\usepackage{amsmath}
\usepackage{bm}
\usepackage{datetime}
\usepackage{braket}
\usepackage[noabbrev]{cleveref}
\usepackage[toc,page]{appendix}
 
\usepackage{titlesec}
\usepackage{setspace}

\usepackage{feynman_diagrams}

\crefrangeformat{equation}{(#3#1#4)--(#5#2#6)}
\numberwithin{equation}{section}

\renewcommand{\L}{\mathcal{L}}
\newcommand{\G}{\mathcal{G}}
\newcommand{\hf}{\frac{1}{2}}
\newcommand{\fb}[2]{\left(\frac{#1}{#2}\right)}
\newcommand{\twovec}[2]{\left(\begin{array}{c}#1\\#2\end{array}\right)}
\newcommand{\abs}[1]{\left|#1\right|}

\newcommand{\msbar}{$\overline{\mathrm{MS}}$ }

\newcommand\blfootnote[1]{%
  \begingroup
  \renewcommand\thefootnote{}\footnote{#1}%
  \addtocounter{footnote}{-1}%
  \endgroup
}

\DeclareMathOperator{\Tr}{Tr}

\newdateformat{monthyeardate}{%
  \monthname[\THEMONTH] \THEYEAR}

\begin{document}

\addtocontents{toc}{\protect\setstretch{1.1}}
\hphantom\\
\begin{flushright}
MAN/HEP/2019/007\blfootnote{\hspace{-1em}$^*$kieran.finn@manchester.ac.uk,\qquad $^\dagger$s.karamitsos@lancaster.ac.uk,\qquad$^\ddagger$apostolos.pilaftsis@manchester.ac.uk}
\\
\monthyeardate{\today}
\end{flushright}

 \title{{\LARGE Frame Covariance in Quantum Gravity}\\ \vspace{1em} }

\author{\large Kieran Finn$^{1\,*}$}
\author{\large Sotirios Karamitsos$^{1,2\,\dagger}$}
\author{\large Apostolos Pilaftsis$^{1\,\ddagger}$}

\affiliation{\vspace{0.4em}$^1$Department of Physics and Astronomy, University of Manchester, Manchester
 M13 9PL, United Kingdom\\
$^2$Consortium for Fundamental Physics, Physics Department, Lancaster University, Lancaster LA1 4YB, United Kingdom}

\begin{abstract}
We develop a quantum effective action for scalar-tensor theories of gravity which is both spacetime diffeomorphism invariant and field reparameterisation (frame) invariant beyond the classical approximation. We achieve this by extending the Vilkovisky-DeWitt formalism, treating both the scalar fields and the components of the gravitational tensor field as coordinates describing a manifold.  By using tensors covariant under diffeomorphisms of this manifold, we show that scalar-tensor theories can be written in a form that is manifestly frame invariant at both classical and quantum levels. In~the same context, we show that in order to maintain manifest frame invariance, we must modify the Feynman rules of theories with a non-trivial field space. We show that one such theory is General Relativity by demonstrating explicitly that it has a non-zero field-space Riemann tensor. Thus, when constructing theories of quantum gravity, we must deal not only with curved spacetime, but also with a curved field space.  Finally, we address the cosmological frame problem by tracing its origin to the existence of a new model function that appears in the path integral measure. Once this function is fixed, we find that frame transformations have no effect on the quantisation of the theory. The uniqueness of our improved quantum effective action is discussed.
\end{abstract}

\maketitle
\flushbottom
\tableofcontents
\section{Introduction}
\label{sec:intro}

The laws of nature should not depend on the way we choose to describe them. While there may be many different ways of parametrising the underlying degrees of freedom in a theory, its physical predictions should not depend on which parametrisation one uses. This seemingly obvious fact has historically had far-reaching consequences. For example, imposing that the laws of physics not care about the way we label space and time leads inevitably to Einstein's celebrated theory of relativity~\cite{Einstein:1915ca}. This idea is known as \emph{reparametrisation invariance} and throughout this paper we use it as a guiding light with the goal of developing a formalism in which reparametrisation invariance is made manifest.

When writing down a Quantum Field Theory (QFT), we must define a set of quantum fields in which to express it. We are always free to re-express the same theory in terms of a different set of fields. This is known as a \emph{change of frame}. There has been much debate in the literature~\cite{Fierz:1956zz,Barvinsky:1985an,Capozziello:1996xg,Faraoni:1998qx,Alvarez:2001qj,Capozziello:2010sc,Steinwachs:2013tr,Jarv:2014hma,Postma:2014vaa,Kamenshchik:2014waa,Domenech:2015qoa,Jarv:2016sow,Herrero-Valea:2016jzz,Pandey:2016unk,Karam:2018squ,Falls:2018olk,Nandi:2019xve,Francfort:2019ynz} as to whether such a change of frame represents an observable change to the theory or merely a change of description.

Since changes of frame correspond to field reparameterisations we expect that they should not affect any physical observables. However, in the ordinary formulation of QFTs, off-shell calculations of quantum corrections can yield different results depending on the set of fields used to perform them, as shown in Appendix~\ref{sec:complex scalar}. Furthermore, it has been shown that when gravity is included, one can get different predictions even for on-shell observables depending on whether the quantum effects are applied before or after changing frame. This has become known as the cosmological frame problem. For a historical overview of the issue, see~\cite{Karamitsos:2018lur}. 

In light of the above issues, our aim is to develop a formalism in which reparametrisation invariance is made manifest both on and off shell and hence does not suffer from the cosmological frame problem. It is important to note that whether or not a formalism is reparametrisation invariant is a consequence of its representation, not its content. Any physical observable of the theory must be invariant under reparametrisations, but this fact can often be obscured by the way the theory is written down.

We emphasise that because the content of a theory has no bearing on reparametrisation invariance, this formalism places no restrictions on which theories are allowed. Therefore reparametrisation invariance cannot be considered a symmetry in the traditional sense and there will, in general, be no Noether current and no gauge degrees of freedom associated with it. 

A better point of comparison is the use of spacetime tensors to highlight the diffeomorphism invariance of a QFT. Although the physical predictions of any theory will necessarily be independent of the spacetime coordinates used to perform the calculation, the use of covariant objects makes this fact manifest. 

In this paper, we shall focus on scalar-tensor theories of quantum gravity~\cite{Jordan1938,jordan1952schwerkraft,Brans:1961sx,Bergmann:1968ve,Wagoner:1970vr,Fujii:2003pa,Faraoni:2004pi} with a field content that consists of a spin-2 graviton field~$g_{\mu\nu}$ and a set of scalar fields $\phi^A$ (collectively denoted as $\bm{\phi}$)
and with an action of the form
\begin{align}
\label{eq:scalar-tensor action}
\begin{aligned}
S&\equiv \int d^4x\sqrt{-g}\, \L\ \, ,&\L &\equiv -\frac{f(\bm{\phi})}{2}R 
 +\hf g^{\mu\nu}k_{AB}(\bm{\phi})\partial_\mu\phi^A\partial_\nu \phi^B-V(\bm{\phi}) ,
\end{aligned}
\end{align}
where $g\equiv\det(g_{\mu\nu})$. Here $f(\bm{\phi})$, $k_{AB}(\bm{\phi})$ and $V(\bm{\phi})$ are the effective Planck mass, the scalar field-space metric and the potential, respectively. We shall refer to these three functions as \emph{model functions} and together they fully define our theory at the classical level. In the context of such theories, there are two types of transformations that amount to nothing more than a change of description --- spacetime diffeomorphisms and field reparameterisations. We wish to write our theory in way that is manifestly invariant under both of these.

Spacetime diffeomorphisms consist of changing the coordinates of spacetime:
\begin{equation}
x^\mu\rightarrow \tilde{x}^\mu=\tilde{x}^\mu(x^\mu).\label{eq:diffeo}
\end{equation}
This is just a relabelling of the points on the spacetime manifold and thus should not affect any physical observables. Diffeomorphism invariance is the backbone of General Relativity and, as such, has been much studied in the literature. We will therefore not focus on it here.

Field reparameterisations involve changing the definition of the fields of the theory by making the transformation
\begin{equation}
\begin{alignedat}{3}
&g_{\mu\nu}&&\to\tilde{g}_{\mu\nu} &&=\tilde{g}_{\mu\nu}(g_{\rho\sigma},\bm{\phi}),\\
&\phi^A&&\to \tilde{\phi}^A &&=\tilde{\phi}^A(g_{\rho\sigma},\bm{\phi}).
\end{alignedat}\label{eq:general reparam}
\end{equation}
Again, this is just a relabelling of the degrees of freedom in the theory and should not have a physical effect.

Spacetime diffeomorphism invariance restricts the class of field redefinitions that we have to consider. When performing the transformation~\eqref{eq:general reparam}, we must maintain the spacetime covariant structure of the fields and should not introduce any new spacetime tensors. If we also insist that our field redefinitions do not mix derivative and non-derivative terms then this restricts the admissible set of transformations to those of the form
\begin{alignat}{3}
&g_{\mu\nu}&&\to \tilde{g}_{\mu\nu}  &&=  \Omega^2(\bm{\phi})g_{\mu\nu},\label{eq:conf trans}\\
& \phi^A&&\to \tilde{\phi}^A   &&=  \tilde{\phi}^A(\bm{\phi}).\label{eq:field redef}
\end{alignat}
We will refer to these transformations as a conformal transformation and a scalar field reparameterisation, respectively. Together, they constitute a \emph{frame transformation}. Under such a frame transformation, the model functions in~\eqref{eq:scalar-tensor action} transform as~\cite{Karamitsos:2017elm}
\begin{alignat}{2}
f&\to\tilde f  &&=\,\Omega^{-2}\,f \;, \nonumber \\
V&\to\widetilde V &&=\,\Omega^{-4}\,V\;,\label{eq:kfv trans}\\
k_{AB}&\to{\tilde k}_{  A   B}&&=  K^C_{\   A} K^D_{\ B} \big[k_{CD}   - 6   f  (\ln \Omega)_{,C} (\ln \Omega)_{,D} + 3  f_{,C} (\ln \Omega)_{,D} +3  (\ln \Omega)_{,C}  f_{,D} \big] \nonumber \;,
\end{alignat}
where a comma $_{,A}\equiv \partial/\partial\phi^A$ denotes differentiation with respect to the field $\phi^A$ and ${K^A_{\  B}\equiv \partial \phi^A/\partial\tilde{\phi}^B}$ is the Jacobian of the scalar field reparameterisation.

In this paper, we will show explicitly, how we can write down a theory in a manifestly reparametrisation invariant way by using the well-known technique of field-space covariance~\cite{Vilkovisky:1984st,GrootNibbelink:2000vx,GrootNibbelink:2001qt,vanTent:2003mn,Burns:2016ric}, whose relevance to resolving the cosmological frame problem was first pointed out in~\cite{Steinwachs:2013ama}. We treat both the scalar fields and the components of the graviton field as coordinates describing a manifold. Frame transformations of the form~\eqref{eq:general reparam} are then simply diffeomorphisms of this manifold. Provided we write down our theory in terms of objects that are both spacetime and field-space tensors, and then fully contract any indices, the theory will be manifestly reparametrisation invariant.

With the field space covariant technique, the theory of General Relativity (which is just a scalar-tensor theory without the scalars) can also be expressed in terms of a field space manifold. This manifold is separate from the spacetime manifold and comes with its own Riemann tensor, Ricci tensor and Ricci scalar. As we will see in Section~\ref{sec:GR}, all these curvature invariants are non-zero. Thus, when studying quantum theories of gravity, we must necessarily deal not only with curved spacetime, but with a curved field space as well. We believe that this observation will be important to consider when constructing a UV complete theory of quantum gravity and may be part of the reason why such a construction has proven so difficult.

We shall express our reparametrisation invariant theory using the quantum effective action~\cite{Sauter:1931zz,Heisenberg:1935qt,Weisskopf:1996bu,Schwinger:1951nm}. All predictions of the theory can be obtained from this effective action and thus defining it is sufficient to fully define the quantum theory. However, as we shall see in Section~\ref{sec:conventional eff action}, the ordinary construction of the effective action depends on our choice of parametrisation.

If we can treat gravity as a classical background, the Vilkovisky-DeWitt (VDW) formalism~\cite{Vilkovisky:1984un,DeWitt:1985sg}, reviewed in Section~\ref{sec:vdw eff action}, is enough to solve this problem. However, if we wish to treat gravity as a field and place it on the same footing as the other fields in our theory, we encounter ambiguities, which inevitably lead to the cosmological frame problem.

As we shall show, these ambiguities arise from a frame-dependent choice that must be made in the standard approach to scalar-tensor theories of gravity. The graviton field~$g_{\mu\nu}$ is normally identified as the metric of spacetime. However, $g_{\mu\nu}$ transforms under a field reparametrisation~\eqref{eq:general reparam} whereas the metric of spacetime does not. This identification is therefore only valid in a particular frame~\cite{Falls:2018olk} and thus the frame invariance of the VDW formalism is ruined.

In this paper we overcome the cosmological frame problem by defining the metric of spacetime in a frame invariant manner. We achieve this through the introduction of a new model function, $\ell = \ell(\bm{\phi})$, so that the metric of spacetime is given by $\bar{g}_{\mu\nu}= g_{\mu\nu}/\ell^2$. We are therefore able to construct, for the first time, a manifestly frame and spacetime diffeomorphism invariant quantum effective action for scalar-tensor theories of gravity.

In practice, the quantum effects of a theory are usually calculated using Feynman diagrams. However, as we shall see in Section~\ref{sec:cov feyn rules}, the usual way in which these diagrams are calculated crucially depends on the frame in which they are evaluated. Feynman rules, when calculated in the usual way, are not covariant field-space tensors and thus different parameterisations of the fields will yield different sets of rules. We will show how the Feynman rules must be modified in the presence of a non-trivial field space. 
 
We adopt the following conventions throughout this paper. Lowercase Greek letters ($\mu$,~$\nu$ etc.) will be used for spacetime indices and repeated indices will imply summation in accordance with the Einstein summation convention. Upper case Latin letters ($A$,~$B$ etc.) will be used for field-space indices with repeated indices again implying summation. Lowercase Latin letters ($a$, $b$ etc.) will be used for configuration-space indices and will thus simultaneously represent both a discrete field-space index and a point in spacetime. For such indices we shall use the Einstein--DeWitt notation~\cite{DeWitt:1967ub} in which repeated configuration-space indices imply summation over the discrete index and integration over spacetime, e.g.
\begin{equation}
J_a\phi^a\equiv\sum_A\,\int d^D x_A \sqrt{-\bar{g}}\, J_A(x_A)\phi^A(x_A), 
\end{equation} 	
where $D$ is the number of spacetime dimensions and $\bar{g}_{\mu\nu}$, with determinant $\bar{g}$, is the metric of spacetime.

This paper is laid out as follows. We begin in Section~\ref{sec:scalars} by reviewing the construction of the field and configuration spaces for scalar field theories. We then review the effective action formalism in Section~\ref{sec:conventional eff action}, explicitly demonstrating that it is dependent on the parametrisation of the fields. We show in Section~\ref{sec:vdw eff action} how Vilkovisky and DeWitt's reformulated effective action resolves these issues when gravity can be treated as a background. In Section~\ref{sec:cov feyn rules}, we show the effect of reparameterisations on ordinary quantum calculations using Feynman diagrams and develop a method for calculating Feynman rules in a reparameterisation-invariant manner.

We show how the same geometric approach of Vilkovisky and DeWitt can be applied to gravity in Section~\ref{sec:GR}, explicitly constructing the field space for General Relativity and showing that this field space is positively curved. We add scalar fields to the theory in Section~\ref{sec:spacetime}, showing that when we do, there is an ambiguity in the definition of the spacetime metric, which is responsible for the cosmological frame problem. In Section~\ref{sec:grand field space}, we construct a field space for the scalar and tensor fields, which we call the grand field space, and use it to write down scalar-tensor theories in a way that is manifestly invariant under a frame transformation~\eqref{eq:general reparam}. We then incorporate the spacetime dependence of the fields  in order to construct a grand configuration space in Section~\ref{sec:grand config space}. This allows us to construct a fully frame and spacetime diffeomorphism invariant path integral measure, which we can then use to quantise the theory in a reparametrisation invariant way. We provide a concise description of the formalism in Section~\ref{sec:summary}, before discussing our findings in Section~\ref{sec:discussion}.

\section{Covariance in Scalar Field Theories}
\label{sec:scalars}

Let us begin by reviewing the construction of the field space for scalar field theories without gravity. Such theories have actions of the form
\begin{align}
\begin{aligned}
S&\equiv \int d^D x\sqrt{-g}\L  , \\
\L &\equiv  \frac{1}{2}g^{\mu\nu}k_{AB}(\bm{\phi})\partial_\mu\phi^A\partial_\nu\phi^B-V(\bm{\phi}) ,
\end{aligned} \label{eq:scalar action}
\end{align}
where $D$ is the dimension of spacetime. In this section we will take the metric of spacetime~$g_{\mu\nu}$ to be fixed and will not consider any redefinitions of the form~\eqref{eq:conf trans}. We will relax this assumption in Section~\ref{sec:GR}.

As discussed in the Introduction, we could just as easily describe this theory in terms of a different set of fields $\tilde{\bm{\phi}}$ and the transformation
\begin{equation}
\phi^A\to\tilde{\phi}^A=\tilde{\phi}^A(\bm{\phi})\label{eq:scalar field redef}
\end{equation}
is just a change of description and should therefore not affect any calculations. In order to make this fact explicit, we will construct a manifold known as the \emph{field space}~\cite{Vilkovisky:1984st,GrootNibbelink:2000vx,GrootNibbelink:2001qt,vanTent:2003mn,Burns:2016ric,Karamitsos:2017elm} and treat the fields $\bm{\phi}$ as coordinates describing that manifold. With such a construction, the transformation~\eqref{eq:scalar field redef} is simply a diffeomorphism of the field space. We can then write down the theory in a way that is explicitly reparameterisation invariant by simply building it out of field-space covariant objects.

The field space is a Riemannian manifold, and so we can equip it with a metric. Such a metric should satisfy the following three properties~\cite{Vilkovisky:1984un}:
\begin{enumerate}
\item It should transform as a symmetric rank 2 tensor under~\eqref{eq:scalar field redef}.
\item It should be determined from the classical action~\eqref{eq:scalar action}.
\item It should be Euclidean for a canonically normalised theory.
\end{enumerate}
The only quantity that satisfies these conditions is the model function $k_{AB}(\bm{\phi})$ and so that is what is used in the literature.

In this paper we want to introduce a new expression for the field-space metric; one that is constructive, rather than relying on the identification of a particular term in Lagrangian. We will thus define the field-space metric to be
\begin{equation}
G_{AB}\equiv \frac{ g_{\mu\nu}}{D}\frac{\partial^2 \L}{\partial(\partial_\mu\phi^A)\partial(\partial_\nu\phi^B)},\label{eq:scalar field space metric}
\end{equation}
where $D$ is the number of spacetime dimensions. Notice that for the theory described by~\eqref{eq:scalar action}, this new prescription still gives ${G_{AB}=k_{AB}}$. However, this new prescription is now constructive and can thus be applied to any field theory --- even, for example, those with higher derivative terms.\footnote{In the case of a higher derivative theory,~\eqref{eq:scalar field space metric} would lead to a Finslerian metric~\cite{Finsler} --- one that depends on both the fields and their derivatives. We will not discuss such theories here, but will save them for future work.} This constructive prescription also ensures that the field space metric is unique for a given theory.

With the field-space metric thus defined, we can straightforwardly define a connection on the field-space manifold:
\begin{equation}
\Gamma^A_{BC}\equiv\hf G^{AD}\left[\frac{\partial G_{BD}}{\partial \phi^C}+\frac{\partial G_{DC}}{\partial \phi^B}-\frac{\partial G_{BC}}{\partial \phi^D}\right]\,,
\label{eq:scalar field space connection}
\end{equation}
where $G^{AB}$ is the inverse of $G_{AB}$. We can also define a field-space covariant derivative
\begin{equation}
\label{eq:scalar field space cov der}
\begin{aligned}
\nabla_C X^A &\equiv\frac{\partial X^A}{\partial \phi^C}+\Gamma^A_{CD}X^D, &\nabla_C X_A &\equiv\frac{\partial X_A}{\partial \phi^C}-\Gamma^D_{CA}X_D,                
\end{aligned}
\end{equation}
and so on in the usual manner for higher-rank tensors.

When quantising the theory, the field-space manifold alone is not sufficient. In the path integral formalism, we must integrate not just over the fields, but over all configurations of the fields. In order to construct this integral in a covariant manner, we define an infinite dimensional configuration-space manifold. Each direction on this manifold represents a different configuration of the fields and thus we can describe it using coordinates
\begin{equation}
\phi^a\equiv \phi^A(x_A).
\end{equation}
The lowercase Latin index $a=\{A,x_A\}$ is a continuous index that runs over all points in spacetime in addition to all the scalar fields in the theory, as described in the Introduction.

In order to define a  metric for the configuration space, we need to add one more property to the list above. The configuration-space metric should be ultra-local, i.e.~it should be proportional to a Dirac delta function only and contain no derivatives of the fields. We therefore define the configuration-space metric as
\begin{align}
\label{eq:scalar config metric}
\begin{aligned}
\G_{ab} \; &\equiv  \; \frac{g_{\mu\nu}}{D}  \frac{\delta^2S}{\delta(\partial_\mu\phi^a)\delta(\partial_\nu\phi^b)} 
 \\
 &= \; G_{AB}\delta^{(D)} (x_A-x_B).   
\end{aligned}
\end{align}
Here we have defined the functional derivative with respect to a partial derivative as
\begin{equation}
\frac{\delta F[\partial_\mu\Phi^A(x)]}{\delta(\partial_\mu\Phi^A(y))} \equiv \lim_{\epsilon^A_\mu\to0}\frac{F[\partial_\mu\Phi^A(x)+\epsilon^A_\mu\delta^{(D)}(x-y)]-F[\partial_\mu\Phi^A(x)]}{\epsilon^A_\mu}, \label{eq:def func deriv}
\end{equation}
where the Dirac delta function is normalised such that
\begin{equation}
\int d^Dx \sqrt{-g}\: \delta^{(D)}(x)=1.
\end{equation}
Such a definition allows $\delta^{(D)}(x)$ to be diffeomorphism invariant.  We note that with this definition $\delta \Phi^A/\delta(\partial_\mu\Phi^A)=0$.

The connection on the configuration-space manifold is as follows:
\begin{align}
\begin{aligned}
\Gamma^a_{bc}&\equiv\hf \G^{ad}\left[\frac{\delta \G_{bd}}{\delta \phi^c}+\frac{\delta \G_{dc}}{\delta \phi^b}-\frac{\delta \G_{bc}}{\delta \phi^d}\right] \\
&=\Gamma^A_{BC}\delta^{(D)}(x_A-x_B)\delta^{(D)}(x_A-x_C),
\end{aligned}\label{eq:scalar config space connection}
\end{align}
and thus the configuration-space covariant functional derivative is
\begin{align}
\begin{aligned}
\nabla_c X^a& \equiv \frac{\delta X^a}{\delta \phi^c}+\Gamma^a_{cd}X^d, &\nabla_c X_a& \equiv \frac{\delta X_a}{\delta \phi^c}-\Gamma^d_{ca}X_d,
\end{aligned}\label{eq:scalar config space cov der}
\end{align}
similar to \eqref{eq:scalar field space cov der}.

With the configuration-space manifold defined, it is straightforward to write theories in a manifestly reparameterisation invariant way. We simply need to build our theory out of configuration-space tensors and ensure that all indices are fully contracted. 

It is also easy to identify quantities that are not reparameterisation invariant.
Two examples of non-invariant objects are the quantum effective action and Feynman diagrams, as we shall show in the following sections.

\section{Non-Covariance of the Ordinary Effective Action}
\label{sec:conventional eff action}

The ordinary effective action formalism~\cite{Sauter:1931zz,Heisenberg:1935qt,Weisskopf:1996bu,Schwinger:1951nm} fundamentally stems from the one-particle irreducible (1PI) approach in QFT. Through its application, it is possible to define an action that inherently incorporates all quantum effects beyond tree level, in principle allowing us to study radiative corrections non-perturbatively.

The starting point for the derivation of the effective action is the generating functional for 1PI diagrams,
\begin{equation}
\label{eq:wdef}
Z[\bm{J}]\equiv \exp\left(\frac{i}{\hbar} W[\bm{J}] \right)=\ \int [\mathcal{D}\bm{\phi}]\; \mathcal{M} [\bm{\phi}]\; \exp \left[\frac{i}{\hbar} S[   \bm{\phi}]  \; +   \; J_a \phi^a \right],
\end{equation}
defined in the presence of an external source field ${J_a\equiv J_A(x_A)}$ (also collectively denoted as~$\bm{J}$). Here the functional integral element is $[\mathcal{D}\bm{\phi}] \equiv \prod_{x,A}d\phi^A(x)$ and~$\mathcal{M} [\bm{\phi}]$ is the measure of the configuration space for the quantum fields $\phi^a$. We have also re-introduced the reduced Planck constant $\hbar$ as a means of keeping track different orders of quantum loops. The generating functional is reminiscent of the partition function in statistical mechanics, which is is a weighted sum of Boltzmann factors over the different microstates of the system. In a similar vein, the generating functional is defined as a weighted integral over all possible configurations of the quantum fields $\phi^a$ of the system.

From the generating functional, it is possible to arrive at the effective action via the Legendre transformation
\begin{align}
\label{eq:gammadef}
\Gamma[\bm{\varphi}] \;  =  \; W[\bm{J}] \;  +i\hbar\;  J_a \varphi^a,
\end{align}
where the $\varphi^a$ (collectively denoted as $\bm\varphi$) are the  \emph{mean fields} and $J_a = J_a[\bm{\varphi}]$ is considered to be a functional of $\bm{\varphi}$. In the presence of the source terms $J_a$, the mean fields and the sources are related by
\begin{align}
\label{eq:meandef}
\varphi^a  =     -i\hbar \frac{\delta W[\bm J]}{\delta J_a},
\qquad
J_a    =     -\frac{i}{\hbar}\frac{\delta \Gamma[\bm{\varphi}]}{\delta \varphi^a}.
\end{align}
The usefulness of the effective action is thus that extremising it generates the quantum-corrected equations of motion.

Already at this point, it is possible to observe that this construction lacks covariance. Since~$\varphi^a$ is not a configuration-space vector, $J_a \varphi^a$ is a frame-dependent expression and thus all three equations~\crefrange{eq:wdef}{eq:meandef} are sensitive to the way in which we parametrise the fields in our theory. This is a major drawback for this approach and one we shall return to, but for now, let us proceed in order to illustrate how the ordinary effective action is usually derived and pave the way for the derivation of the covariant expression. 
 
The effective action $\Gamma[\bm{\varphi}]$ satisfies the following implicit functional integro-differential equation: 
\begin{equation}
\label{eq:standard effective action}
\exp\left(\frac{i}{\hbar} \Gamma[\bm{\varphi}]  \right)=\ \int [\mathcal{D}\bm{\phi}]\, \mathcal{M} [\bm{\phi}]\, \exp \left\{\frac{i}{\hbar}  \Big[ S[   \bm{\phi}] + \frac{\delta\Gamma[\bm{\varphi}]}{\delta\varphi^a} (\varphi^a - \phi^a ) \Big]\right\}. \nonumber
\end{equation}
Equation~\eqref{eq:standard effective action} may be derived by substituting~\eqref{eq:gammadef} and~\eqref{eq:meandef} in~\eqref{eq:wdef}.
Evidently, solving~\eqref{eq:standard effective action} exactly is prohibitively hard. Fortunately, it is possible to solve for~$\Gamma[\bm{\varphi}]$ in a perturbative
loop-wise expansion with the help of the \emph{background field method}~\cite{Abbott:1981ke}, where we split the quantum field $\phi^a$ into a background component, which we treat classically, and a quantum perturbation. 
Similarly, we expand ${\Gamma[\bm{\varphi}] = S_0[\bm{\varphi}] +  \hbar \Gamma^{(1)}[\bm{\varphi}] + \hbar^2  \Gamma^{(2)}[\bm{\varphi}] + \cdots}$. At~each loop order, the path integral can be evaluated explicitly. In detail, at one and two-loop order,
we have
 \begin{align}
 \label{eq:standard effective action1loop}
\Gamma^{(1)} [  \bm{\varphi}]   &=  \;   i\ln \mathcal{M} [  \bm{\varphi}]   \; -  \;  \frac{i}{2} \ln  {\rm det}  S_{,ab}[\bm{\varphi}] \;,\\
\Gamma^{(2)}  [  \bm{\varphi}]   \; &= \;  \frac{1}{8}\Delta^{ab}\Delta^{cd}S_{,abcd}-\frac{1}{12}\Delta^{ab}\Delta^{cd}\Delta^{ef}S_{,ace}S_{,bdf},\label{eq:eff act 2 loop}
 \end{align}
where a comma $_{,a}\equiv \delta/\delta\phi^a$ indicates a functional derivative with respect to the field $\phi^a$ and
\begin{equation}
\Delta^{ab}\equiv\fb{\delta^2 S}{\delta\phi^a\delta\phi^b}^{-1}
\end{equation}
is the propagator.

As we shall explore in detail in Section~\ref{sec:cov feyn rules},~\eqref{eq:eff act 2 loop} can be represented graphically by the Feynman diagrams
\begin{equation}
\label{eq:gamma2}
\Gamma^{(2)}  [  \bm{\varphi}]   \; = \;\SolidInfinity+\SolidBasketball \ .
\end{equation}
Note that $\Gamma^{(2)}   [  \bm{\varphi}]  $ contains only 1PI graphs. Other possible one-particle reducible diagrams, such as
\begin{equation}
\SolidDumbbell
\end{equation}
evaluate to zero and so do not contribute to the final expression \eqref{eq:gamma2}.

For our theory to be fully reparametrisation invariant, we require that the effective action be a scalar under reparameterisations of the mean fields
\begin{align}\label{eq:mean frame trans}
\varphi^a  \; &\rightarrow  \; \widetilde\varphi^a=\widetilde\varphi^a(\bm{\varphi}).
 \end{align}
We saw above that the explicit dependence of the generating functional on the fields $\varphi^a$ spoils the covariance, and as a result, such a transformation will not leave \eqref{eq:standard effective action1loop} and~\eqref{eq:eff act 2 loop}  invariant. 
This occurs because the difference $\varphi^a - \phi^a$ does not transform as a vector in configuration space, spoiling the covariance of the term $\frac{\delta\Gamma[\bm{\varphi}]}{\delta\varphi^a} (\varphi^a - \phi^a )$ in~\eqref{eq:standard effective action}. Similarly, the presence of ordinary functional derivatives in~\eqref{eq:standard effective action1loop} and~\eqref{eq:eff act 2 loop} induces extra terms in the expression for $\Gamma^{(1)}$ and~$\Gamma^{(2)}$, which means that the expression for the effective action is not a configuration-space scalar. The parametrisation dependence of the effective action can be seen explicitly in Appendix~\ref{sec:complex scalar}.

\section{Vilkovisky and DeWitt's Solution: The Covariant Effective Action} 
\label{sec:vdw eff action}

The Vilkovisky-DeWitt (VDW) effective action formalism~\cite{Vilkovisky:1984un,Vilkovisky:1984st,DeWitt:1985sg,Rebhan:1986wp,Huggins:1986ht,Kunstatter:1986qa,Kunstatter:1990hz} was developed in order to address the problems of non-covariance of the ordinary effective action that were outlined in the previous section. Unlike the conventional approach, this formalism does not unduly privilege a particular frame. In this section, we review the key results of the VDW formalism.

As noted in Section~\ref{sec:conventional eff action}, the non-invariance of the ordinary effective action stems from the term $\frac{\delta\Gamma}{\delta\varphi^a}(\phi^a-\varphi^a)$ in~\eqref{eq:standard effective action}, which is not a configuration-space scalar. Vilkovisky's proposal~\cite{Vilkovisky:1984un} was therefore to replace the difference $\phi^a-\varphi^a$ with a two-point quantity $\Sigma^a [\bm{\varphi},\bm{\phi}]$ that transforms as a vector with respect to the mean field $\bm{\varphi}$, a scalar with respect to the quantum field $\bm{\phi}$ and satisfies $\Sigma^a[\bm{\phi},\bm{\phi}]=0$. Making this replacement in~\eqref{eq:standard effective action} gives
\begin{equation}
   \label{eq:VDW effective action}
 {\exp\left(\frac{i}{\hbar}   \Gamma[\bm{\varphi}]  \right)} =\ \int [\mathcal{D}\bm{\phi}]\, \mathcal{M} [\bm{\phi}]\, \exp \left\{ \frac{i}{\hbar}
\Big[ S[   \bm{\phi}] \; +   \; \frac{\delta\Gamma[\bm{\varphi}  ]}{\delta\varphi^a} \, \Sigma^a[\bm{\varphi},\bm{\phi}]   \,   \Big]  \right\}.
\end{equation}
There are no frame-dependent terms in~\eqref{eq:VDW effective action} and therefore this newly defined action is fully frame invariant.

Vilkovisky's original proposal was to use ${\Sigma^a[\bm{\varphi},\bm{\phi}]=\sigma^a[\bm{\varphi},\bm{\phi}]}$, where $\sigma^a[\bm{\varphi},\bm{\phi}]$ is the tangent vector to the geodesic connecting $\bm{\varphi}$ and~$\bm{\phi}$ evaluated at~$\bm{\varphi}$. The affinely normalised tangent vector can be found by solving
\begin{align}
\label{eq:sdef1}
 \sigma^b[\bm{\varphi},\bm{\phi}]\,\nabla_b \sigma^a  [\bm{\varphi},\bm{\phi}] \;=  \; \sigma^a [\bm{\varphi},\bm{\phi}],
\end{align}
along with the boundary conditions
  \begin{align}
  \label{eq:sdef2}
  \begin{aligned}
  \sigma^a  [\bm{\varphi},\bm{\phi}] \big\vert_{\bm{\varphi} = \bm{\phi}} \; &= \; 0,\\
 \nabla_b \sigma^a  [\bm{\varphi},\bm{\phi}]\big\vert_{\bm{\varphi} = \bm{\phi}} \;  &=  \; \delta^A_B\delta^{(D)}(x_A-x_B)\equiv\delta^a_b,
 \end{aligned}
\end{align}
where $\nabla_a$ is the covariant derivative as defined in~\eqref{eq:scalar config space cov der} and is taken to act on the first argument $\bm{\varphi}$. It is possible to expand $\sigma^a[\bm{\varphi},\bm{\phi}]$ in terms of the configuration-space connection~$\Gamma^a_{bc}[\bm{\varphi}]$ as
\begin{equation}
 -\sigma^a[\bm{\varphi},\bm{\phi}]  =  -(\varphi^a-\phi^a) + \frac{1}{2}\, \Gamma^a_{bc}[\bm{\varphi}] (\varphi^b-\phi^b)(\varphi^c-\phi^c)  + \cdots.
\end{equation}

However, $\sigma^a$ is not the only possible choice of two-point quantity that satisfies the required properties to make the action frame invariant. In fact, any superposition of tangent vectors
 \begin{align}
\label{eq:def Sigma}
\Sigma^a [\bm{\varphi},\bm{\phi}] =  (C^{-1}[\bm{\varphi}])^a_{\ b} \, \sigma^b[\bm{\varphi},\bm{\phi}]
\end{align}
will do. We therefore need to introduce another requirement to fix the matrix $C^a_b$.

For theories with a flat configuration space we can always go to a frame in which the metric is Euclidean and all the connections vanish. In such a frame there should be no non-trivial field-space effects and thus the VDW effective action should agree with the ordinary effective action calculated in the previous section. It can be shown~\cite{Ellicott:1987ir} that this requirement forces us to choose $C^a_b=\delta^a_b$ for such theories. However, for theories with non-zero configuration-space curvature, no such frame exists and so a different condition is required to fix $C^a_b$.

The choice made by DeWitt~\cite{DeWitt:1985sg} is the condition of vanishing tadpoles
\begin{equation}
\langle\Sigma^a[\bm\varphi,\bm\phi]\rangle_{\Sigma}=0,\label{eq:Sigma condition}
\end{equation}
where the expectation value is defined as
\begin{equation}
\langle F[\bm\varphi,\bm\phi]\rangle_{\Sigma}={\exp\left(-\frac{i}{\hbar}   \Gamma[\bm{\varphi}]  \right)}
\int [\mathcal{D}\bm{\phi}]\, \mathcal{M} [\bm{\phi}]\, F[\bm\varphi,\bm\phi] \exp \left\{ \frac{i}{\hbar}
\Big[ S[   \bm{\phi}]  +  \frac{\delta\Gamma[\bm{\varphi}  ]}{\delta\varphi^a} \, \Sigma^a[\bm{\varphi},\bm{\phi}]  \,   \Big]\right\}.
\end{equation} 
This choice was made for two main reasons. First, it allows the effective action to be calculated perturbatively as a sum of 1PI Feynman diagrams~\cite{Burgess:1987zi}. Second, when the formalism is extended to gauge theories,~\eqref{eq:Sigma condition} is vital in ensuring that the resulting effective action is independent of the choice of gauge-fixing conditions~\cite{Rebhan:1986wp,Ellicott:1987ir}.

In order to satisfy~\eqref{eq:Sigma condition}, we find that we require~\cite{DeWitt:1985sg}
\begin{equation}
 \label{eq:expand C}
C ^a_{\ b}[\bm\varphi]=\langle \nabla_b\sigma^a[\bm\varphi,\bm\phi]\rangle_{\Sigma}= \langle \delta^a_b - \frac{1}{3}  R^a_{\ cbd}[\bm{\varphi}]\ \sigma^c [\bm{\varphi}, \bm{\phi}]\; \sigma^d[\bm{\varphi}, \bm{\phi}] + \ldots\rangle_{\Sigma}\,. 
\end{equation}
Here $R^a_{\ cbd}$ is the Riemann tensor of the configuration-space manifold. Notice that the Riemann tensor for a flat manifold is $R^a_{cbd}=0$ and thus we recover Vilkovisky's original proposal in this case.
 
We can use the background field method to expand~\eqref{eq:VDW effective action} perturbatively, exactly as we did for the ordinary effective action~\eqref{eq:standard effective action}. This gives us the following equations for the one and two-loop corrections to the VDW effective action~\cite{Ellicott:1987ir}
 \begin{align} 
\Gamma^{(1)}  [  \bm{\varphi}]   \; = \;  &-\frac{i}{2}\ln \det   \nabla^a \nabla_b S, \label{eq:vdw 1 loop}\\
\Gamma^{(2)}  [  \bm{\varphi}]   \; = \;  &\frac{1}{8}\Delta^{ab}\Delta^{cd}\nabla_{(a}\nabla_b\nabla_c\nabla_{d)} S \label{eq:vdw 2 loop}-\frac{1}{12}\Delta^{ab}\Delta^{cd}\Delta^{ef} \big(\nabla_{(a}\nabla_c\nabla_{e)} S\big)\big(\nabla_{(b}\nabla_d\nabla_{f)} S\big) ,
 \end{align}
where $\Delta^{ab}=\left(\nabla_a\nabla_b S\right)^{-1}$ is the covariant propagator and the parentheses $(\dots)$ denote symmetrisation with respect to the indices enclosed. Notice that $\Gamma^{(1)}$ and $\Gamma^{(2)}$ are both now invariant under a frame transformation~\eqref{eq:mean frame trans} as expected.

It was noted in~\cite{Ellicott:1987ir,Burgess:1987zi} that the VDW effective action defined in~\eqref{eq:VDW effective action} does not generate the covariant correlation functions of $\bm\varphi$ in its current form. In order to achieve this, we must instead define 
\begin{align}\label{eq:base point effective action}
 {\exp\left(\frac{i}{\hbar}   \widetilde{\Gamma}[\bm{\varphi},\bm{\varphi}_0]  \right)}=\ \int [\mathcal{D}\bm{\phi}]\, \mathcal{M} [\bm{\phi}]\, \exp \left\{ \frac{i}{\hbar}
\Big[ S[   \bm{\phi}] \; +   \; \frac{\delta\widetilde{\Gamma}[\bm{\varphi},\bm{\varphi}_0 ]}{\delta\varphi^a} \,\big( \Sigma^a[\bm{\varphi}_0,\bm{\phi}]-\Sigma^a[\bm{\varphi}_0,\bm{\varphi}]\big)   \,   \Big] \right\},
\end{align}
where $\bm{\varphi}_0$ is an arbitrary base point.

The effective action in~\eqref{eq:base point effective action} depends explicitly on the base point $\bm{\varphi}_0$, and so one may question its uniqueness. However, as shown in~\cite{Burgess:1987zi}, this \emph{explicit} dependence of $\widetilde{\Gamma}[\bm{\varphi},\bm{\varphi}_0]$ on $\bm{\varphi}_0$ gets cancelled against the \emph{implicit} dependence of $\bm{\varphi} = \bm{\varphi} (\bm{\varphi}_0)$ evaluated at the same base point,~i.e.
\begin{equation}
  \label{eq:BKaction}
\frac{\delta}{\delta\varphi_0^a}\, 
\widetilde{\Gamma}[\bm{\varphi}(\bm{\varphi}_0),\bm{\varphi}_0]\ =\ 0\,.
\end{equation}
Hence, $\widetilde{\Gamma}[\bm{\varphi}(\bm{\varphi}_0),\bm{\varphi}_0]$ is independent of~$\bm{\varphi}_0$.

As a consequence, one may consider a simplified scheme, in which~$\bm{\varphi}_0$ is identified with~$\bm{\varphi}$, such that $\Sigma^a[\bm{\varphi}_0,\bm{\varphi}]$ vanishes on the RHS of~\eqref{eq:base point effective action}. 
In this simplified scheme, we recover  the VDW effective action, where $\Gamma[\bm \varphi ]=\widetilde{\Gamma}[\bm\varphi,\bm\varphi]$. However, when calculating 
higher order $n$-point correlation functions,  the above identification of $\bm{\varphi}_0$ with
$\bm{\varphi}$ must be made \emph{only after} any covariant differentiation with respect to $\bm{\varphi}$ in order to avoid introducing spurious terms. 

For brevity, we shall only present the VDW effective action~\eqref{eq:VDW effective action} in this paper. Nevertheless, it is straightforward to introduce a base point and generalise to~\eqref{eq:base point effective action} by simply making the replacement
\begin{equation}
\Sigma^a[\bm{\varphi},\bm{\phi}]\ \to\ \Sigma^a[\bm{\varphi}_0,\bm{\phi}] - \Sigma^a[\bm{\varphi}_0,\bm{\varphi}]\,. 
\end{equation}

It is important to note that, on shell, we have $\frac{\delta\Gamma}{\delta\varphi^a}=0$, in which case the expressions for the ordinary effective action~\eqref{eq:standard effective action} and the VDW effective action~\eqref{eq:VDW effective action} are identical. Thus, we are guaranteed to get the same results for on-shell observables regardless of whether we use the ordinary effective action~\eqref{eq:standard effective action} or the VDW effective action~\eqref{eq:VDW effective action}. This also means that any parametrisation dependence that arises when using the ordinary effective action must vanish when the calculations are performed on shell. We show some examples of this in Appendix~\ref{sec:complex scalar}. 

The fact that the VDW formalism remains covariant off-shell is important for a few reasons, even if off-shell quantities will never appear in observables. First, from a geometric point of view, we expect covariance to be satisfied for the entirety of the configuration space, not just the geodesics. The ordinary approach is parametrisation-independent only for a severely restricted subspace (the on-shell region), and so the VDW approach is required to restore covariance for the whole configuration space. Second, off-shell formulations of QFTs have many important applications, such as in supersymmetry~\cite{Sohnius:1985qm} and the analysis of quantum anomalies~\cite{Grigore:2010ug}. Finally, inflationary observables are often computed in the slow-roll approximation~\cite{Liddle:1994dx}. Such an approximation forces us to perform calculations in the off-shell regime.

\section{Covariant Feynman Rules}
\label{sec:cov feyn rules}

In the previous section, we showed how the quantum effective action can be constructed in a fully covariant way. However, in practice, radiative corrections are often calculated perturbatively with the help of Feynman diagrams. As we will show in this section, usual Feynman diagrams are also inherently non-covariant. As such, their form depends on the parametrisation used to calculate them. They should therefore be replaced with an alternative, fully covariant method of calculating Feynman rules. Such a covariant expansion was first developed by Honerkamp~\cite{Honerkamp:1971sh,Ecker:1973qf,Honerkamp:1974bp} in the context of chiral pion theories, but can be readily extended to any scalar field theory as shown below. In this section we provide an explicit derivation of the formalism before applying it to specific examples in appendices~\ref{sec:complex scalar},~\ref{sec:curved field}, and~\ref{sec:non renorm}.

We first review how ordinary Feynman diagrams may be employed to calculate correlation functions (as well as S-matrix elements through the LSZ reduction formula \cite{Lehmann:1954rq}). The derivation can be found in most textbooks on QFT (see, e.g.,~\cite{Pokorski2000QFT}) , but our treatment here most closely follows~\cite{srednicki2007quantum} and~\cite{ryder1996quantum}.

In the path integral formulation of QFT, a correlation function in the presence of a source~$\bm J$, is given by
\begin{equation}
\left\langle\phi^a \phi^b \ldots ,\bm{J}\right\rangle=\frac{\int \left[{\cal D}\bm{\phi}\right]\mathcal{M} [\bm{\phi}](\phi^a \phi^b \ldots )e^{\frac{i}{\hbar} S[\bm{\phi_0+\phi}]+J_a \phi^a}}{\int \left[{\cal D}\bm{\phi}\right]\mathcal{M} [\bm{\phi}]e^{\frac{i}{\hbar} S[\bm{\phi_0+\phi}]+J_a\phi^a  }},\label{eq:define correlation func}
\end{equation}
where $\bm{\phi_0}$ is an arbitrary point around which we quantise~-- usually taken to be the classical vacuum. This can be calculated using the generating functional $Z[\bm{J}]$ defined in~\eqref{eq:wdef}. In terms of this generating functional, the correlation function becomes
\begin{equation}
\left\langle\phi^{a}\phi^{b}\ldots ,\bm{J}\right\rangle=\frac{1}{Z[\bm{J}]}\left(\frac{\delta}{\delta J_{a}}\frac{\delta}{\delta J_{b}}\ldots \right)Z[\bm{J}].\label{eq:correlation Z}
\end{equation}

In order to perform perturbative calculations, we use a Taylor series expansion of the action
\begin{equation}
S[\bm{\phi_0+\phi}]=\sum_N  S^{(N)}_{{a_1}\ldots {a_N}}\phi^{a_1} \ldots \phi^{a_N}\label{eq:standard taylor}
\end{equation}
where
\begin{equation}
S^{(N)}_{{a_1} \ldots {a_N}}=\frac{1}{N!}\left.\frac{\delta^N S}{\delta  \phi^{a_1} \ldots \delta\phi^{a_N}}\right|_{\phi_0}.\label{eq:standard taylor term}
\end{equation}
The constant term $S^{(0)}$ gives factors which cancel out in~\eqref{eq:define correlation func} and therefore we will therefore ignore it. We will also take $\bm{\phi_0}$ to be the classical vacuum so that we have $S^{(1)}_a=0$.  The lowest order non-trivial term in our expansion is therefore
\begin{equation}
S[\bm{\phi_0+\phi}]\approx  S^{(2)}_{ab}\, \phi^a\phi^b,
\end{equation}
about which we shall expand the generating functional $Z[\bm{J}]$.

We must also Taylor expand the path integral measure. Using Vilkovisky's suggestion of $\mathcal{M}[\bm\phi]=\sqrt{\det \G_{ab}}$ we find find~\cite{Vilkovisky:1984st}
\begin{align}
\mathcal{M}[ \bm{\varphi_0}\!+\! \bm{\varphi}]\! = \!1\! + \delta^{(D)}(0)\!\!\int\!\! d^Dx\sqrt{-g} \Tr\ln G_{AB}[\bm\phi(x)]\!+\ldots.
\end{align}
From this expansion, we see that all non-trivial effects of the measure are proportional to~$\delta^{(D)}(0)$. This is a simple divergence equal to the total volume of the spacetime manifold and, as such will be removed by our regularisation procedure. Therefore, the functional form of the measure will have no impact on perturbative results and so we can set $\mathcal{M}=1$. 

With this knowledge, and the expansion given in \eqref{eq:standard taylor term}, we can write
\begin{equation}
Z[\bm{J}]=\exp\left(\frac{i}{\hbar} \sum_{N>2} S^{(N)}_{{a_1} \ldots {a_N}} \frac{\delta}{\delta J_{a_1} } \ldots \frac{\delta}{\delta J_{a_N}}\right)\int \left[{\cal D}\bm{\phi}\right]e^{ \frac{i}{\hbar^2} S^{(2)}_{ab}\phi^a\phi^b+   J_a\phi^a  },
\end{equation}
The functional integral is now Gaussian and so can be calculated explicitly. The result is
\begin{equation}
Z[\bm{J}]={\cal N}\exp\left( \frac{i}{\hbar} \sum_{N>2} S^{(N)}_{{a_1} \ldots {a_N}} \frac{\delta}{\delta J_{a_1}} \ldots \frac{\delta}{\delta J_{a_N}}\right) \exp\left( -i\hbar J_a\Delta^{ab} J_b\right)\,,
\end{equation}
where $\Delta^{ab}$ is the inverse of $S^{(2)}_{ab}$, often known as the propagator, and $\cal N$ is an irrelevant normalisation factor.

Expanding out the two exponentials, we see that the correlation function~\eqref{eq:correlation Z} is
\begin{align}
&\left\langle\phi^a\phi^b\ldots ,\bm{J}\right\rangle\label{eq:correlation func expanded}\\
&=\frac{{\cal N}}{Z[\bm{J}]}\left(\frac{\delta}{\delta J_a}\frac{\delta}{\delta J_b}\ldots \right)\prod_{N>2}\left[\sum_{V_N=0}^{ \infty}\frac{1}{V_N!}\left(\frac{i}{\hbar}S^{(N)}_{{a_1}\ldots {a_N}} \frac{\delta}{\delta J_{a_1}}\ldots \frac{\delta}{\delta J_{a_N}}\right)^{V_N}\right]
\sum_{P=0}^\infty\frac{1}{P!}\left(-i\hbar J_c\Delta^{cd} J_d\right)^P.\nonumber
\end{align}

Feynman diagrams~\cite{Feynman1949ab} are a beautiful graphical way to keep track of the non-zero terms in~\eqref{eq:correlation func expanded}. If we represent each propagator by a line,
\begin{equation}
\LabelledSolidProp{a}{b}=\Delta^{ab},\label{eq:standard prop}
\end{equation}
and each term of the expansion~\eqref{eq:standard taylor} with a vertex,
\begin{equation}
\SolidNPoint{a_1}{a_2}{a_3}{a_N}=N! \, S^{(N)}_{{a_1}\ldots {a_N}}=\left.\frac{\delta^N S}{\delta \phi^{a_1}\ldots \delta\phi^{a_N}}\right|_{\bm\phi_0}\,,
\label{eq:standard feyn rule}
\end{equation}
then each term in~\eqref{eq:correlation func expanded} can be expressed as a diagram with $P$ propagators and $V_N$ vertices of order $N$. Calculating the correlation function then simply amounts to summing up all possible diagrams with the correct number of external legs. Finally, it can easily be shown that the prefactor $\frac{\cal N}{Z[\bm{J}]}$ on the RHS of~\eqref{eq:correlation func expanded} has the effect of removing all diagrams that are not fully connected.

The above derivation is very elegant and has been used extensively in QFT  calculations. However, it is not reparametrisation invariant. This is because, as we have seen, the quantity $\phi^a$ is not a configuration-space vector. Therefore, it will not transform in a covariant manner and cannot be contracted to form reparameterisation invariant quantities.

This means that the individual terms on the RHS of~\eqref{eq:standard taylor} will change under a field redefinition. Although the full sum will remain invariant (since the LHS \emph{is} a configuration-space scalar), the individual terms will mix into each other and hence any finite truncation of the sum will not be invariant. Moreover, the term $J_a\phi^a$ in~\eqref{eq:wdef}, as well as the definition of the correlation function~\eqref{eq:define correlation func} are not field covariant. As such, their form is dependent on our choice of parametrisation.  Some examples of the parametrisation dependence of ordinary Feynman calculations are shown in Appendix~\ref{sec:complex scalar}.

It is therefore clear that a new, covariant approach to Feynman diagrams is required if we are to calculate quantum corrections in a fully covariant manner. The simplest way to achieve such invariance is to replace the coordinate $\phi^a$ with a configuration-space vector, much like we did in Section~\ref{sec:vdw eff action}. However, in contrast to the previous section, we will employ Vilkovisky's original choice and choose it to be the tangent vector in configuration space $\sigma^a[\bm{\phi_0,\phi_0+\phi}]$. We shall therefore calculate the covariant correlation functions
\begin{equation}
\left\langle\sigma^a\sigma^b\ldots ,\bm{J}\right\rangle_\sigma=\frac{\int \left[\cal{D}\bm{\sigma}\right](\sigma^a\sigma^b\ldots )e^{\frac{i}{\hbar}S[\bm{\phi_0+\phi}]+J_a\sigma^a }}{\int \left[\cal{D}\bm{\sigma}\right]e^{\frac{i}{\hbar}S[\bm{\phi_0+\phi}]+J_a\sigma^a }},\label{eq:covariant correlation function}
\end{equation}
where the suppressed arguments of $\sigma^a$ are $[\bm{\phi_0,\phi_0+\phi}]$ in all cases. Notice that $\left[{\cal D}\bm{\phi}\right]\mathcal{M} [\bm{\phi}]=\left[\cal{D}\bm{\sigma}\right]$ and thus the measure is trivial in this case.

We note that $\sigma^a[\bm{\phi_0,\phi_0+\phi}]=\phi^a+O(\phi^2)$ and therefore the correlation functions~\eqref{eq:define correlation func} and~\eqref{eq:covariant correlation function} have the same pole structure. This means that the renormalised on-shell S matrix elements
\begin{equation}
\prod_{I=1}^E\lim_{k_I^2\to m_I^2}\frac{k_I^2-m_I^2}{Z_I^\hf}\left\langle\sigma^a(k_1)\sigma^b(k_2)\ldots ,0\right\rangle_\sigma =\prod_{I=1}^E\lim_{k_I^2\to m_I^2}\frac{k_I^2-m_I^2}{Z_I^\hf}\left\langle\phi^a(k_1)\phi^b(k_2)\ldots ,0\right\rangle\label{eq:S matrix}
\end{equation}
are identical~\cite{Lee:1975vt}. Here $E$ is the number of external fields in the correlation function and~$m_I$ and $Z_I$ are the (renormalised) mass and wavefunction renormalisation of particle $I$, respectively. Off shell, however, the correlation functions~\eqref{eq:define correlation func} and~\eqref{eq:covariant correlation function} will not be equal in general.

Note that we should continue to use the correlation functions~\eqref{eq:covariant correlation function} to calculate S-matrix elements even in the presence of field-space curvature. The correlation functions of DeWitt's modified two-point quantity~$\bm\Sigma$ give only linear combinations of~\eqref{eq:S matrix}, as can be seen from~\eqref{eq:def Sigma}, and therefore should not be used.

Let us modify the definition of the generating function to make it frame invariant:
\begin{equation}
\widetilde{Z}[\bm{J}]=\int \left[{\cal D}\bm{\sigma}\right]e^{\frac{i}{\hbar} S[\bm{\phi_0+\phi}]+ J_a\sigma^a[\bm{\phi_0,\phi_0+\phi}]}. 
\end{equation}
We then find that the correlation functions are given by
\begin{equation}
\left\langle\sigma^a\sigma^b\ldots ,\bm{J}\right\rangle_\sigma=\frac{1}{\widetilde{Z}[\bm{J}]}\left(\frac{\delta}{\delta J_a}\frac{\delta}{\delta J_b}\ldots \right)\widetilde{Z}[\bm{J}].\label{eq:cov correlation Z}
\end{equation}
Finally, we consider an alternative, but equivalent, covariant expansion of the action~\cite{Vilkovisky:1984un}, given by
\begin{equation}
S[\bm{\phi_0+\phi}]=\sum_N{\widetilde S}^{(N)}_{a_1...a_n}\sigma^{a_1}[\bm{\phi_0,\phi_0+\phi}]\ldots \sigma^{a_n} [\bm{\phi_0,\phi_0+\phi}]\,,
\label{eq:covariant taylor}
\end{equation}
where
\begin{equation}
{\widetilde S}^{(N)}_{a_1 \ldots a_n}=\frac{1}{N!}\left.\nabla_{(a_1} \ldots \nabla_{a_n)} S\right|_{\bm\phi_0},\label{eq:covariant taylor term}
\end{equation}
and $(\cdots)$ refers to symmetrisation over all indices. Now, since $\sigma^a$ is a genuine field-space vector, all ${\widetilde S}^{(N)}$  are fully covariant field-space tensors and every term in~\eqref{eq:covariant taylor} is independently reparameterisation invariant.

We can repeat the same derivation as above to calculate the correlation functions graphically by using Feynman diagrams. Now, however, the Feynman rules must be calculated covariantly with the propagator being given by
\begin{equation}
\LabelledSolidProp{a}{b}=(\left.\nabla_a\nabla_b S\right|_{\bm{\phi}_0})^{-1},
\end{equation}
and the vertex factors given by
\begin{equation}
\SolidNPoint{a_1}{a_2}{a_3}{a_n}=\left.\nabla_{(a_{1}}\ldots \nabla_{a_n)}S\right|_{\bm{\phi}_0}.\label{eq:cov feyn rule}
\end{equation}

Notice that the Feynman rule is symmetrised over its indices. This is because only the symmetrised version of~\eqref{eq:covariant taylor term} appears in~\eqref{eq:covariant taylor}. For~\eqref{eq:standard feyn rule}, this symmetrisation had no effect since the ordinary functional derivative is already symmetric. However, for theories with curved field space, covariant functional derivatives do not commute and as a result, this symmetrisation is vital in fixing the order of differentiation. 

In Appendices~\ref{sec:complex scalar},~\ref{sec:curved field}, and~\ref{sec:non renorm}, we perform some explicit calculations using the covariant Feynman approach, demonstrating its relation to results obtained in the ordinary approach.

\section{The Geometric Structure of Gravity}
\label{sec:GR}

So far we have treated gravity as a background and have not considered the metric $g_{\mu\nu}$ to be a field. However, the Vilkovisky-DeWitt covariant approach explored in the previous sections can be readily applied to tensor fields~\cite{Vilkovisky:1984un,Falls:2018olk,Bounakis:2017fkv}. We therefore promote $g_{\mu\nu}$ to become a fully dynamical field. Doing so will lead us to the construction of the field space for gravitational theories. This space is a Riemannian manifold and is distinct from the manifold of spacetime. The goal of this section is to illustrate the geometrical features of gravity as described by General Relativity.

We begin by examining the action for General Relativity, described by the Einstein--Hilbert action,
 \begin{align}
S = -\frac{1}{2}\int d^D x \, \sqrt{-g} \, R\,. \label{eq:EH action}
\end{align}
We use the standard definitions
 \begin{align}
 \begin{alignedat}{2}
& \Gamma^{\alpha}_{\mu\nu}&&= \frac{g^{\alpha\beta}}{2} \big( g_{\beta \nu, \mu } + g_{\mu \beta, \nu } -g_{\mu\nu,\beta}  \big),\\
&R^\alpha{}_{\mu\beta\nu} &&=  \Gamma^\alpha_{\nu\mu,\beta}
    -  \Gamma^\alpha_{\beta\mu,\nu}
    + \Gamma^\alpha_{\beta\lambda}\Gamma^\lambda_{\nu\mu}
    - \Gamma^\alpha_{\nu\lambda}\Gamma^\lambda_{\beta\mu},\\
 &R _{\mu \nu} &&=  R^\alpha{}_{\mu\alpha\nu},\\
 &R  &&=   g^{\mu\nu} R _{\mu \nu},
  \end{alignedat}
\end{align}
for the spacetime Christoffel symbols, Riemann tensor, Ricci tensor and Ricci scalar, respectively.

This action is, famously, nonrenormalizable~\cite{tHooft:1974toh,Goroff:1985sz}. Because of this, the construction of a UV-complete quantum theory of gravity has yet to be achieved and remains one of the most important open problems in physics. Performing such a construction is far beyond the scope of this paper and, as such, we make no attempt to solve the issue of nonrenormalizability. However, we believe that the issues identified in this section, in particular the curvature of the field space of GR, will be important to consider in any future worked aimed at solving these problems.

For $D$-dimensional gravity, the $D(D+1)/2$ degrees of freedom of the field space will be represented by an unordered pair of spacetime indices $(\mu\nu)$.
In order to maintain consistency with the position of the indices, we take the fundamental field to be $g^{\mu\nu}$. This means that~ $\delta g^{\mu\nu}$ is a contravariant vector in field space and $\delta g_{\mu\nu}$ is a covariant vector.

As discussed in Section~\ref{sec:scalars}, the field-space metric can be explicitly calculated from the classical action~\eqref{eq:EH action} by using~\eqref{eq:scalar field space metric}. However, there is a subtlety with gravity, stemming from its gauge freedom. This freedom requires us to add to the action a gauge fixing term of the form
\begin{align}
\label{eq:gauge fixing}
S_{\rm GF} =  -\frac{\gamma}{2} \int d^D x \, \sqrt{-g}\;\chi^\mu g_{\mu\nu }\chi^\nu.
\end{align}
Here, $\chi^\mu=0$ is the gauge fixing condition and $\gamma$ is a non-negative constant. When we apply~\eqref{eq:scalar field space metric} to the sum of~\eqref{eq:EH action} and~\eqref{eq:gauge fixing}, we get the metric\footnote{Note that we can also arrive at~\eqref{eq:general gr metric} up to an irrelevant normalisation simply by enforcing that $G_{(\mu\nu)(\rho\sigma)}$ transforms as a spacetime tensor and is symmetric under $\mu\leftrightarrow\nu$ and $\rho\leftrightarrow\sigma$.}
\begin{equation}
G_{(\mu\nu)(\rho\sigma)}=\frac{1}{2} \left( g_{\mu\rho} g_{\sigma\nu } +   g_{\mu \sigma } g_{\rho \nu}  - \alpha g_{\mu\nu} g_{\rho\sigma} \right),\label{eq:general gr metric}
\end{equation}
where $\alpha=\alpha(\chi^\mu, \gamma)$ is a constant that depends on the gauge fixing condition $\chi^\mu$ and the constant~$\gamma$. For example, in de Donder gauge $g^{\rho\sigma}\Gamma^\mu_{\rho\sigma}=0$, we have $\alpha=2-\gamma$.

Because $\chi^\mu$ and $\gamma$ are both arbitrary, we need another condition to fix $\alpha$. The condition we choose is
\begin{equation}
\left(G^{-1}\right)^{(\mu\nu)(\rho\sigma)}=\;G^{(\mu\nu)(\rho\sigma)}\equiv\; g^{\alpha \mu} g^{\beta \nu} g^{\kappa \rho} g^{\lambda \sigma}\, G_{(\alpha  \beta) \,  (\kappa \lambda)},\label{eq:gr fix alpha condition}
\end{equation}
where $\left(G^{-1}\right)^{(\mu\nu)(\rho\sigma)}$ is the inverse metric satisfying
\begin{equation}
\label{eq:inverse GR metric}
G_{(\mu\nu)(\rho\sigma)}\left(G^{-1}\right)^{(\rho\sigma) \, (\kappa \lambda)}\, =\, \frac{1}{2} (\delta^\mu_\rho \delta^\nu_\sigma + \delta^\mu_\sigma \delta^\nu_\rho) .
\end{equation}
Mathematically, this condition is useful, since it means that there is no difference between raising $(\mu\nu)$ indices with the spacetime metric or the field space metric.

The inverse metric can be calculated from~\eqref{eq:inverse GR metric} and is found to be
\begin{align}
 \label{eq:inverse gravity metric}
\left(G^{-1}\right)^{(\mu\nu) \, (\rho\sigma)} &= \frac{1}{2} \left( g^{\mu\rho} g^{\nu\sigma}+g^{\mu\sigma} g^{\rho\nu}-  \frac{2 \alpha}{ D \alpha  - 2} g^{\mu\nu} g^{\rho\sigma} \right).
\end{align}
The solution to~\eqref{eq:gr fix alpha condition} is therefore\footnote{The solution $\alpha=0$ also allows satisfies this equation. However, we choose to use the solution in~\eqref{eq:alpha}, since it agrees with Vilkovisky's original calculation~\cite{Vilkovisky:1984un} in $D=4$, as well as other results in the literature~\cite{Odintsov:1991yx,Giulini:1994dx}.}
 \begin{align}
\alpha = \frac{4}{D}.\label{eq:alpha}
\end{align}
Thus, in four dimensions,~\eqref{eq:general gr metric} reduces to
\begin{align}
G_{(\mu\nu)(\rho\sigma)}=P_{\mu\nu\rho\sigma}\equiv \frac{1}{2} \left( g_{\mu\rho} g_{\sigma\nu } +   g_{\mu \sigma } g_{\rho \nu}  -  g_{\mu\nu} g_{\rho\sigma} \right),\label{eq:vilk metric}
\end{align}
where $P_{\mu\nu\rho\sigma}$ is Vilkovisky's metric for gravity, derived in~\cite{Vilkovisky:1984un} by different considerations.

Note that this differs from the DeWitt metric~\cite{DeWitt:1967yk}, which imposes a time slicing condition and focuses only on the spatial part of the spacetime metric. In contrast, our calculation considers all components of the spacetime metric equally. This allows the metric~\eqref{eq:vilk metric} to transform as a tensor under diffeomorphisms of the full spacetime.

We note that the metric~\eqref{eq:vilk metric} can be projected onto the space of gauge orbits if one wants to maintain manifest gauge invariance of the VDW effective action~\cite{Vilkovisky:1984un}. While gauge dependence of the effective action is an important topic, and indeed was one of the original motivations for Vilkovisky's work, it runs parallel to our objective of frame invariance and has been much studied in the literature~\cite{DeWitt:1980jv,Vilkovisky:1984st,Rebhan:1986wp,Fradkin:1983nw,Kunstatter:1986qa}. For simplicity we shall therefore ignore this complication in the remainder of the paper. The configuration space metric~\eqref{eq:vilk metric} is good enough to achieve our goal of manifest reparametrisation invariance.

We are now equipped to determine the curvature of the field space for gravity. The expressions for the curvature tensors are identical to those for spacetime, but with spacetime indices replaced with field-space indices. Thus, the field-space Christoffel symbols and Riemann tensor are given as
\begin{align} 
\begin{aligned} 
\Gamma^{(\alpha\beta)}_{(\mu\nu) \, (\rho\sigma)} &= \hf P^{\alpha\beta \, \gamma\delta}\Big(  \partial_{(\mu\nu)}P_{\gamma\delta  \, \rho\sigma}   + \partial_{(\rho\sigma)}P_{\mu\nu \,   \gamma\delta}   - \partial_{(\gamma\delta)}P_{\mu\nu \, \rho\sigma}  \Big),
\end{aligned}
\end{align}
\begin{align} 
\begin{aligned}
\mathfrak{R}^{(\mu\nu)}{}_{(\alpha\beta)(\rho\sigma)(\gamma\delta)} &= 
\partial_{(\rho\sigma)}\Gamma^{(\mu\nu)}_{(\gamma\delta)(\alpha\beta)}
    - \partial_{(\gamma\delta)}\Gamma^{(\mu\nu)}_{(\rho\sigma)(\alpha\beta)}+ \Gamma^{(\mu\nu)}_{(\rho\sigma)(\kappa\lambda)}\Gamma^{(\kappa\lambda)}_{(\gamma\delta)(\alpha\beta)}
    - \Gamma^{(\mu\nu)}_{(\gamma\delta)(\kappa\lambda)}\Gamma^{(\kappa\lambda)}_{(\rho\sigma)(\alpha\beta)} ,
\end{aligned}
\end{align}
respectively, where $\partial_{(\mu\nu)} \equiv \partial/\partial g^{\mu\nu}$. Correspondingly, the field-space Ricci tensor and Ricci scalar for gravity are given by
\begin{align} 
\begin{aligned}
\mathfrak{R}_{(\alpha\beta) (\gamma\delta)}& =  \mathfrak{R}^{(\mu\nu)}{}_{(\alpha\beta)(\mu\nu)(\gamma\delta)},&\mathfrak{R}  &=   P^{\alpha\beta \, \gamma\delta} \, \mathfrak{R}_{(\alpha\beta)  (\gamma\delta)}\,.
\end{aligned}
\end{align}

To cope with the complexity of this calculation, we employed the symbolic computer algebra system
{\tt Cadabra2}~\cite{Peeters:2007wn, cadabra2}. In this way we find the following explicit forms for the Riemann tensor $\mathfrak{R}^{(\mu\nu)}{}_{(\alpha\beta)(\rho\sigma)(\gamma\delta)}$ (shown in Appendix~\ref{sec:configuration riemann}), the Ricci tensor
\begin{equation}
\mathfrak{R}_{(\mu\nu)( \rho\sigma)} = \frac{1}{4}g_{\mu\nu} g_{\rho\sigma} - \frac{D}{8} g_{\mu\rho} g_{\nu\sigma} - \frac{D}{8} g_{\mu\sigma} g_{\nu\rho},
\end{equation}
and the Ricci scalar
\begin{equation}
\mathfrak{R} = \frac{D}{4}- \frac{{D}^{2}}{8} - \frac{{D}^{3} }{8}.\label{eq:gr ricci}
\end{equation}
These tensors are all non-zero (except when $D=1$, which is expected since one dimensional curvature is impossible). Therefore this shows that gravity has a genuinely curved field space. Indeed, we can see from~\eqref{eq:gr ricci} that the field space is always negatively curved. It would be interesting to explore whether this negative curvature is the origin for the non-convergence of the path integral for pure gravity.

\section{The Cosmological Frame Problem in Scalar-Tensor \mbox{Theories}}
\label{sec:spacetime}

After studying scalar field theories and gravitational theories separately, we now wish to combine the methods of the previous sections and look at theories with both scalar fields and gravity. In the following two sections, we will therefore construct a covariant formalism for scalar-tensor theories with an action of the form~\eqref{eq:scalar-tensor action}.

However, before we do so, we must address the \emph{cosmological frame problem}, which stems from a subtlety regarding spacetime diffeomorphism invariance in scalar-tensor theories. Diffeomorphism invariance is normally achieved by identifying the graviton field~$g_{\mu\nu}$ as the metric of spacetime and thus defining the spacetime line element as
\begin{equation}
ds^2=g_{\mu\nu}dx^\mu dx^\nu.\label{eq:normal spacetime line}
\end{equation}
However, when $g_{\mu\nu}$ is taken to be a dynamical field, the RHS of~\eqref{eq:normal spacetime line} is no longer reparametrisation invariant. Indeed it picks up a conformal factor $\Omega^2$ under a conformal transformation~\eqref{eq:conf trans}. In contrast, the spacetime line element $ds^2$ is a measurable quantity and so must be invariant under reparametrisations of the fields.

Previous authors~\cite{Falls:2018olk,Kamenshchik:2014waa,Ohta:2017trn} have dealt with this (either explicitly or implicitly) by choosing a `preferred frame' in which the frame-dependent relation~\eqref{eq:normal spacetime line} holds. Different choices of this preferred frame lead to different quantum corrections for otherwise identical theories, even when these corrections are calculated on shell.

In order to avoid the cosmological frame problem, we shall use a different, frame invariant, definition of the metric of spacetime. The most general such definition that does not require the introduction of any new spacetime tensors and has no momentum dependence is
\begin{equation}
\bar{g}_{\mu\nu}\equiv\frac{g_{\mu\nu}}{\ell^2(\bm{\phi}(x))}\label{eq:spacetime metric}
\end{equation}
where $\bar{g}_{\mu\nu}$ is the metric of spacetime and $\ell$ is a (generally spacetime dependent) length scale. In this paper, we will restrict ourselves to the case where $\ell$ depends on $x$ only through the scalar fields $\bm\phi$ in which case $\ell(\bm{\phi})$ represents another non-singular model function in our theory.\footnote{Note that if we do not make this assumption, then $\ell(x)$ would act as a new field in the theory and we would have to quantise it accordingly.}  

Provided that $\ell$ transforms as
\begin{equation}
\ell\to\tilde{\ell}=\Omega\, \ell\label{eq:l trans}
\end{equation}
under conformal transformations~\eqref{eq:conf trans} and does not transform under scalar field redefinitions~\eqref{eq:field redef}, then $\bar{g}_{\mu\nu}$ is frame invariant. Thus, we may define a spacetime line element
\begin{equation}
\label{eq:spacetime line element}
d\bar{s}^2=\bar{g}_{\mu\nu}dx^\mu dx^\nu
\end{equation}
which is both frame and diffeomorphism invariant. This line element is also dimensionless, in contrast to the standard definition, and therefore qualifies as an observable according to the Buckingham-$\pi$ theorem~\cite{Buckingham:1914ab}. Previous authors~\cite{Higgs:1959jua,Flanagan:2004bz,Catena:2006bd,Jarv:2014hma} have defined similar frame invariant line elements, but have assumed a particular form of $\ell$. To the best of our knowledge, we are the first to identify~$\ell(\bm\phi)$ as a freely selectable model function that must be specified when defining a scalar-tensor theory.

At first glance, it may appear that $\ell$ has no physical meaning. After all, it does not appear anywhere in the classical action~\eqref{eq:scalar-tensor action}, and so will have no effect on any classical observable. However, as we will show below, $\ell$ \emph{does} appear in the functional measure of the path integral and therefore the choice of $\ell$ will have an observable impact at the quantum level. 

Specifying a particular form of $\ell(\bm\phi)$ is mathematically equivalent to  specifying a ``preferred frame" in the ordinary approach. The frame in which the metric of spacetime $\bar{g}_{\mu\nu}=g_{\mu\nu}$ is the one in which $\ell(\bm\phi)=1$, which we shall refer to as the \emph{metric frame}. The effects of a non-trivial $\ell(\bm\phi)$ are therefore equivalent to the so-called ``frame descrimanent" calculated in~\cite{Falls:2018olk}.

However, in our formalism the metric frame is no more ``preferred" than any other and thus we are able to write down a scalar-tensor theory of gravity without ever singling out a particular frame. This distinction, although subtle, is vital in constructing a unique, reparametrisation invariant effective action. In addition, by defining a reparametrisation invariant spacetime line element~\eqref{eq:spacetime line element} we can see explicitly how the form of $\ell(\bm\phi)$ affects the theory, as we will show in the following sections.

Note that in general the Einstein frame and the metric frame are different, and it is not always possible to choose a frame both with minimal coupling $f(\bm\phi)=1$, and with $\ell(\bm\phi)=1$. Because of this, the metric $\bar{g}_{\mu\nu}$ will not, in general, obey Einstein's equations even in the Einstein frame. This is to be expected. Einstein's equations are the equations of motion that arise upon varying the action with respect to the gravitational tensor field $g_{\mu\nu}$. It is therefore this tensor field that obeys Einstein's equations (in the Einstein frame) and it is $g_{\mu\nu}$ that will form part of our grand field space.

Let us therefore clarify the difference between $\bar{g}_{\mu\nu}$ and~$g_{\mu\nu}$. The metric $\bar{g}_{\mu\nu}$ is the metric of spacetime and therefore appears in the spacetime line element~\eqref{eq:spacetime line element} as well as in the construction of all spacetime invariant objects. Conversely, $g_{\mu\nu}$ is a field, on the same footing as the scalar fields $\bm\phi$, and thus appears in all equations of motion as well as in Feynmann diagrams. When solving the equations of motion to calculate the field configuration, it is $g_{\mu\nu}$ that must be calculated and thus Einstein's equations may still be used (provided one works in the Einstein frame).

\section{The Grand Field Space}
\label{sec:grand field space}

In this section, we will construct an augmented field-space manifold that incorporates both the scalar fields $\phi^A$ and the gravitational tensor field $g^{\mu\nu}$~\cite{Fujikawa:1983im,Falls:2018olk}. To this end, we shall define the following coordinate chart:
\begin{equation}
\Phi^I=\twovec{g^{\mu\nu}}{\phi^A},\label{eq:grand field}
\end{equation}
where $I=\{\mu\nu,A\}$. We call this augmented space the \emph{grand field space}.

As mentioned in the Introduction, any physical observable should be invariant under reparameterisations of the fields. Such reparameterisations are nothing but diffeomorphisms of the grand field space. In fact, the transformations~\eqref{eq:general reparam} can be re-expressed in this notation as
\begin{equation}
\Phi^I\to\widetilde{\Phi}^I(\bm{\Phi}).\label{eq:grand trans}
\end{equation}

We now equip our grand field space with a metric. We wish to define the metric in such a way that both spacetime diffeomorphism invariance and invariance under~\eqref{eq:grand trans} remains manifest. To do so, we first define the invariant Lagrangian
\begin{equation}
\bar{\L}=\ell^D\L
\end{equation}
such that
\begin{equation}
\label{invarlagr}
S=\int d^Dx\sqrt{-g}\L=\int d^Dx \sqrt{-\bar{g}}\, \bar{\L} \, .
\end{equation}
This definition allows $\bar{\L}$ to be invariant under both spacetime diffeomorphisms and~\eqref{eq:grand trans}. This is in contrast to the standard Lagrangian~$\L$, which picks up a conformal factor under the conformal transformation~\eqref{eq:conf trans}.

We can now define the metric of the grand field space in a way analogous to~\eqref{eq:scalar field space metric}. Explicitly, we have
\begin{equation}
\label{eq:GIJ}
G_{IJ}=\frac{\bar{g}_{\mu\nu}}{D}\frac{\partial^2 \bar{\L}}{\partial(\partial_\mu\Phi^I)\partial(\partial_\nu\Phi^J)}.
\end{equation}
It is important to note that the effective Planck length~$\ell$ is now part of the definition of the field-space metric, since it appears in the definition of $\bar{g}_{\mu\nu}$.

For the scalar-tensor theory described by~\eqref{eq:scalar-tensor action} in four dimensions the field space metric is
\begin{equation}
G_{IJ}= \ell^2\begin{pmatrix}
f P_{\mu\nu\rho\sigma}&- \frac{3}{4} f_{,B} g_{\mu\nu}\\
-\frac{3}{4} f_{,A}g_{\rho\sigma}& k_{AB}
\end{pmatrix}\;,\label{eq:grand field space metric}
\end{equation}
where $P_{\mu\nu\rho\sigma}$ is defined in \eqref{eq:vilk metric}.

We note that, as discussed in Section~\ref{sec:GR}, the metric~\eqref{eq:grand field space metric} does not follow directly from~\eqref{eq:GIJ}, unless the gauge fixing term takes a specific form. However, previous works~\cite{DeWitt:1980jv,Vilkovisky:1984st,Rebhan:1986wp,Ellicott:1987ir,Fradkin:1983nw,Kunstatter:1986qa} have shown that one can define a \emph{projected} field-space metric from $G_{IJ}$ in~\eqref{eq:grand field space metric}, such that the resulting VDW effective action is independent of the gauge-fixing condition. One can therefore use the techniques developed in these works in order construct this projected field-space metric~\eqref{eq:grand field space metric} in a unique way.

This metric can be used to define a frame-invariant field-space line element, given by
\begin{equation}
d\sigma^2=G_{IJ}d\Phi^Id\Phi^J.
\end{equation}
By construction, this line element is both spacetime-diffeomorphism invariant and frame invariant. We can also define the connection on the grand field-space as 
\begin{equation}
\Gamma^I_{JK}=\hf G^{IL}\left[\partial_J G_{LK}+\partial_K G_{JL}-\partial_L G_{JK}\right],
\end{equation}
where $\partial_I\equiv \partial/\partial\Phi^I$. The form of the connection can then be used to construct the field-space covariant derivative
\begin{align}
\nabla_J X^I&=\frac{\partial X^I}{\partial \Phi^J}+\Gamma^I_{JK}X^K,&\nabla_J X_I&=\frac{\partial X_I}{\partial \Phi^J}-\Gamma^K_{JI}X_K,
\end{align}
with straightforward generalisation to higher order tensors. Anything constructed out of field-space tensors and the field-space covariant derivative will be invariant under~\eqref{eq:grand trans} provided all indices are properly contracted.

\section{The Grand Configuration Space}
\label{sec:grand config space}
We now wish to extend the geometric construction of the grand field space in order to take into account the spacetime dependence of the fields. This means that each coordinate now comes with a spacetime argument,
\begin{equation}
\Phi^i\equiv\Phi^I(x_I).\label{eq:define config}
\end{equation}
As in section~\ref{sec:scalars}, the lowercase Latin index~$i=\{I,\bm{x}_I\}$ is a continuous index and runs over all points in spacetime as well as all the fields in our theory.

In order to maintain both manifest diffeomorphism and frame invariance, we will make use of the invariant spacetime metric~\eqref{eq:spacetime metric} and define the spacetime line element as in \eqref{eq:spacetime line element}. We will also use the corresponding invariant volume element when performing spacetime integrals and from now on, integrations of repeated configuration space indices will be performed as
\begin{equation}
X_iY^i\equiv \sum_I\,\int d^D x_I \sqrt{-\bar{g}}\, X_I(x_I)Y^I(x_I).
\end{equation}

This choice of spacetime metric directly affects the definition of both the functional derivative and the functional determinant, and we will be explicit in defining them such that their dependence on the metric is made clear.

With the help of the spacetime metric $\bar{g}_{\mu\nu}$, we can define functional differentiation as follows:
\begin{equation}
\frac{\bar{\delta}F[\Phi(x)]}{\bar{\delta}\Phi(y)} \equiv \lim_{\epsilon\to0}\frac{F[\Phi(x)+\epsilon\bar{\delta}^{(D)}(x-y)]-F[\Phi(x)]}{\epsilon}, \label{eq:def bar deriv}
\end{equation}
where we have defined
\begin{equation}
\label{eq:def delta bar}
\bar{\delta}^{(D)}(x) \equiv \ell^D\delta^{(D)}(x)
\end{equation}
such that
\begin{equation}
\int d^Dx \sqrt{-\bar{g}} \, \bar{\delta}^{(D)}(x)=1.\label{eq:delta bar}
\end{equation}
With the definition \eqref{eq:def delta bar}, $\bar{\delta}^{(D)}(x)$ is both diffeomorphism and frame invariant. As a result, functional derivatives defined as in \eqref{eq:def bar deriv} will inherit their transformation properties from the functional~$F$ and field $\Phi$.

Notice that in general $\bar{g}_{\mu\nu}$ depends on all of the grand field space coordinates $\Phi^I$ and therefore so does $\bar{\delta}^{(D)}(x)$. This means that derivatives of the form
\begin{equation}
\frac{\bar{\delta}}{\bar{\delta}\Phi^i}\bar{\delta}^{(D)}(x)\neq0~\label{eq:delta not constant}
\end{equation}
will be non zero. This is not just a consequence of the definition~\eqref{eq:spacetime metric}. Even with the standard non-invariant definitions (i.e. with $g_{\mu\nu}$ identified as the metric) the diffeomorphism invariant Dirac delta function cannot be treated as a constant once the metric is dynamical.

The condition~\eqref{eq:delta not constant} causes the functional derivative~\eqref{eq:def bar deriv} not to commute. This is not a problem. As the calculation in section~\ref{sec:cov feyn rules} shows, Feynman rules must be calculated in a symmetric way and therefore there is no ambiguity stemming from the order of derivatives. Furthermore,~$\bar{\delta}^{(D)}(x)$ has no dependence on $\partial_\mu\Phi^i$ and so the definition of the configuration space metric~\eqref{eq:scalar config metric} can be generalised straightforwardly as shown below.

The choice of metric $\bar g_{\mu\nu}$ also affects how we take the functional determinant, since for an infinite dimensional matrix, the determinant involves an integral over the continuous degrees of freedom. We must therefore explicitly choose which volume measure we will use to count them. Using the invariant volume element derived from~\eqref{eq:spacetime line element}, the functional determinant is given by
\begin{equation}
\label{eq:def bar det}
\overline{\det}(M_{xy}) \equiv \exp\left[i\int d^Dx \sqrt{-\bar{g}} \ln(M)_{xx}\right].
\end{equation}

We have written both the functional derivative and the functional determinant with an overbar to emphasise that these are defined with respect to the metric $\bar{g}_{\mu\nu}$. Using any other metric (e.g. $g_{\mu\nu}$) would lead to a non-equivalent definition and, in general, would not maintain diffeomorphism and frame invariance.

These definitions allow us to define the metric of the grand configuration space as follows:
\begin{equation}
\begin{aligned}
\G_{ij} &\equiv \frac{\bar{g}_{\mu\nu}}{D}\frac{\bar{\delta}^2 S}{\bar{\delta}(\partial_\mu\Phi^i)\bar{\delta}(\partial_\nu\Phi^j)}
\\
&=G_{IJ}(x_I)\bar{\delta}^{(D)}(x_I-x_J).
\end{aligned}\label{eq:config metric}
\end{equation}

The uniqueness of the configuration-space metric was questioned by DeWitt (see discussion in Section 14 of~\cite{DeWitt:1985sg}). Indeed, without the introduction of the model function $\ell$, there would be an ambiguity as to which spacetime metric should be used in the definition~\eqref{eq:config metric}~\cite{Falls:2018olk}. In our prescription, however, $\ell$ is a fundamental part of the theory, no less important than $f$,~ $k_{AB}$ or~$V$. Therefore, for a given theory, $\ell$ must have a fixed functional form and hence the definition~\eqref{eq:config metric} is unique. Nonetheless, as we discuss below, one needs to take care in eliminating any further dependence of the effective action on gauge-fixing conditions.

With the help of the grand configuration space metric, we may write down the line element of the grand configuration-space as 
\begin{equation}
{\cal D}\Sigma^2[\bm{\Phi}]\equiv\G_{ij}{\cal D}\Phi^i{\cal D}\Phi^j=\int d^Dx\sqrt{-\bar{g}} \;  G_{IJ}(x) {\cal D}\Phi^I(x){\cal D}\Phi^J(x) \; .  
\end{equation}
We can also construct the configuration-space connection, given by the Christoffel symbols
\begin{equation}
\Gamma^i_{jk}\equiv\hf \G^{il}\left[\frac{\bar{\delta}\G_{jl}}{\bar{\delta}\Phi^k}+\frac{\bar{\delta}\G_{lk}}{\bar{\delta}\Phi^j}-\frac{\bar{\delta}\G_{jk}}{\bar{\delta}\Phi^l}\right]\label{eq:config christoffels}
\end{equation}
and hence a covariant functional derivative
\begin{align}
\overline{\nabla}_j X^i&=\frac{\bar{\delta} X^i}{\bar{\delta} \Phi^j}+\Gamma^i_{jk}X^k,&\overline{\nabla}_j X_i&=\frac{\bar{\delta} X_i}{\bar{\delta} \Phi^j}-\Gamma^k_{ji}X_k,\label{eq:config cov der}
\end{align}
with straightforward generalisation to higher order tensors.

The invariant configuration-space line element allows us to construct an invariant path-integral volume element 
\begin{equation}
\left[\,\overline{\cal D}\bm{\Phi}\right]\sqrt{\overline{\det}\left(\G_{ij}\right)}.\label{eq:measure}
\end{equation}
Note that the functional integral element~${\left[\,\overline{\cal D}\bm{\Phi}\right]=\prod_{x,I} d\Phi^I(x)}$ is the product of integral elements at every point in spacetime. How these points are counted depends crucially on the metric of spacetime and therefore~$\left[\,\overline{\cal D}\bm{\Phi}\right]$ will depend on the model function~$\ell$. We have highlighted this by denoting it with an overbar to emphasise the choice of $\bar{g}_{\mu\nu}$ as the metric of spacetime.

We can see the dependence explicitly using the identity $\prod_i A_i=\exp\left(\sum_i\ln(A_i)\right)$, which holds for discrete products and can be extrapolated to continuous products. We therefore have
\begin{equation}
\left[\,\overline{\cal D}\bm{\Phi}\right]=\exp\left[\sum_I\int d^Dx\, \sqrt{-\bar{g}(x)}\,\ln\big({\cal D}\Phi^I(x)\big)\right].
\end{equation}

Finally, we must ensure that gauge fixing is done in a reparametrisation invariant manner. We therefore modify the gauge-fixing term~\eqref{eq:gauge fixing} to be
\begin{align}
\label{eq:bar gauge fixing}
S_{\rm GF}[\bm\Phi] =  -\frac{\gamma}{2} \int d^D x \,  \sqrt{-\bar{g}}\; \chi^\mu(\bm\Phi) \bar{g}_{\mu\nu }\chi^\nu(\bm\Phi).
\end{align}
Defined in this way, the gauge fixing condition~$\chi^\mu(\bm\Phi)$ is a grand configuration space scalar.

We note that the gauge fixing condition is, in general, a function not only the tensor field $g_{\mu\nu}$, but also the scalar fields $\bm\phi$. Even if a gauge condition~$\chi^\mu$ that depends only on $g_{\mu\nu}$ is chosen in some frame (for example the De Donder gauge used earlier), it will pick up a dependence on $\bm \phi$ after a frame transformation.

At this stage, one may worry whether the gauge-fixing term in~\eqref{eq:bar gauge fixing} will threaten the uniqueness of~\eqref{eq:config metric}. However, previous works in the literature~\cite{DeWitt:1980jv,Vilkovisky:1984st,Rebhan:1986wp,Ellicott:1987ir,Fradkin:1983nw,Kunstatter:1986qa} have shown that it is possible to define a \emph{projected} configuration space metric leading to an effective action which is independent of the gauge fixing condition. Therefore, one can employ the techniques developed in these earlier works, as well as the model function~$\ell(\bm\phi)$ to define 
such a projected configuration space metric in a well-defined and unique manner.

Gauge fixing also requires us to include the Faddeev--Poppov determinant~\cite{Faddeev:1967fc} in our path integral measure. This can be defined in a frame-invariant way as
\begin{equation}
V_{\mathrm{FP}}=\overline{\det}\left(\frac{\bar{\delta}\chi^\mu(\bm x)}{\bar{\delta} \xi^\nu(\bm y)}\right),
\end{equation}
where $\xi^\mu$ are the gauge parameters. With this term included, we see that the path integral measure is
\begin{equation}
{\cal M}[\bm\Phi]=V_{\mathrm{FP}}\sqrt{\overline{\det}\left(\G_{ij}\right)}.
\end{equation}

We can use the above constructions to define a diffeomorphism and frame invariant effective action:
\begin{equation}
\exp\left(\frac{i}{\hbar}\Gamma[\bm{\varphi}]\right)\label{eq:cov eff act}=\int\! \left[\,\overline{\cal D}\bm{\Phi}\right]{\cal M}[\bm\Phi]\, \exp\bigg[\frac{i}{\hbar}\!\bigg(\!S[\bm{\Phi}]\!+\!\frac{\bar{\delta}\Gamma[\bm{\varphi}]}{\bar{\delta}\varphi^i}\;\Sigma^i[\bm{\varphi},\bm{\Phi}]\! \bigg)\bigg],
\end{equation}
where $\bm{\varphi}=(g^{\mu\nu}\!\!,\;\bm\phi)$ collectively denote the grand fields of~\eqref{eq:grand field} and $S[\bm{\Phi}]$ includes both the classical action~\eqref{eq:scalar-tensor action} and the gauge fixing term~\eqref{eq:bar gauge fixing}. The effective action defined in this way satisfies the important property
\begin{equation}
\Gamma[\bm\varphi;\ell(\bm\phi),f(\bm\phi),k_{AB}(\bm\phi),V(\bm\phi)\,]=\Gamma[\widetilde{\bm\varphi}(\bm\varphi);\tilde{\ell}(\bm\phi),\tilde{f}(\bm\phi),\tilde{k}_{AB}(\bm\phi),\widetilde{V}(\bm\phi)\,]\, .\label{eq:same gamma}
\end{equation}
In~\eqref{eq:same gamma}, the transformations of $\widetilde{\bm\varphi}$, $\tilde{\ell}$ and ($\tilde{f}$, $\tilde{k}_{AB}$, $\widetilde{V}$) are given by~\eqref{eq:grand trans},~\eqref{eq:l trans} and~\eqref{eq:kfv trans}, respectively. Thus, by construction, $\Gamma$ is manifestly frame invariant.

Given that the configuration space metric~$G_{ij}$ depends on the definition of $\bar g_{\mu\nu}$, there will be non-trivial effects of the model function $\ell(\bm\phi)$ at the quantum level arising from the measure~\eqref{eq:measure}.

These effects are equivalent to those calculated in~\cite{Falls:2018olk}. However, here we provide an alternative interpretation. Instead of arising from a mismatch between the frame that we choose to work in and some preferred frame in which the theory is quantised, the `frame discriminant' is a simply the one-loop effects of the model function~$\ell(\bm\phi)$. In our approach, the choice of measure does not introduce an ambiguity in the definition of the effective action. Instead, it is part of the theory itself. Therefore, theories with different measures will have different but unique effective actions.

\section{Summary of the Frame Covariant Formalism}
\label{sec:summary}

In this section, we summarise our frame covariant formalism for scalar-tensor theories. To fully specify a scalar-tensor theory, we require four model functions:
\begin{enumerate}
\item the effective Planck length $\ell$,
\item the effective Planck mass $f$,
\item the scalar field-space metric $k_{AB}$,
\item the scalar potential $V$.
\end{enumerate}
In detail, with these model functions, the classical action is given by
\begin{equation}
S[\bm\Phi]=\int d^Dx \sqrt{-g}\left[-\frac{f R}{2}+ \frac{k_{AB}}{2}\partial_\mu\phi^A\partial^\mu \phi^B-V\right]+S_{\rm GF}[\bm\Phi],
\end{equation}
where $\bm\Phi$ are the grand fields given by~\eqref{eq:grand field}, spacetime indices are contracted with $g_{\mu\nu}$, and $S_{\rm GF}$ is given by~\eqref{eq:bar gauge fixing}.
We can then extract the metric of the grand configuration space,~$\G_{ij},$ from the classical action using
\begin{equation}
\G_{ij}[\bm\Phi]=\frac{\bar{g}_{\mu\nu} }{D} \frac{\bar{\delta}^2 S[\bm\Phi]}{\bar{\delta}(\partial_\mu\Phi_I(x_I))\bar{\delta}(\partial_\nu\Phi_J(x_J))},
\end{equation}
where $\bar{g}_{\mu\nu}= g_{\mu\nu}/\ell^2$ is the metric of spacetime as given in \eqref{eq:spacetime metric}.

We can calculate the quantum effects of this theory in two equivalent ways. One way is to use the VDW action~$\Gamma[\bm{\varphi}]$. This can be calculated from the implicit equation~\eqref{eq:cov eff act}.

An alternative way in which quantum corrections can be calculated is through the use of covariant Feynman diagrams as described in Section~\ref{sec:cov feyn rules}. In this approach, Feynman rules are calculated in a covariant manner with the propagators given by
\begin{equation}
\LabelledSolidProp{i}{j}=\left(\nabla_i\nabla_j S|_{\bm\Phi_0}\right)^{-1}
\end{equation}
and the vertices given by
\begin{equation}
\SolidNPoint{i_1}{i_2}{i_3}{i_n}=\nabla_{(i_1}\ldots \nabla_{i_n)}S|_{\bm\Phi_0},
\end{equation}
where $\bm\Phi_0$ is the base point of the perturbation, usually taken to be the classical vacuum. Feynman diagrams can then be calculated in the usual way.

Both of the above approaches agree with the standard calculation for on-shell observables, but they additionally preserve reparametrisation invariance off shell.

\section{Conclusions}
\label{sec:discussion}

We have developed a covariant formalism for scalar-tensor theories of quantum gravity. By extending the Vilkovisky-DeWitt effective action and the geometric structure of the configuration space, we have constructed a Quantum Field Theory that is manifestly frame and spacetime diffeomorphism invariant.

This is in contrast to previous approaches, which required us to identify a ``preferred frame'' in which the expression for the ordinary effective action~\eqref{eq:standard effective action} holds. The non-covariance of~\eqref{eq:standard effective action} leads to an inequivalence in the standard approach between theories with different choices of preferred frame. This is the root of the cosmological frame problem.

Our formalism resolves this issue by identifying a new model function~$\ell(\bm\phi)$ that relates the spacetime metric~$\bar{g}_{\mu\nu}$ and the gravitational tensor field~$g_{\mu\nu}$. Choosing the form of~$\ell({\bm \phi})$  in our formalism is equivalent to choosing a preferred frame in the conventional approach but does not unduly privilege a particular frame.

In addition, we have seen how the choice of spacetime metric affects the contraction of DeWitt indices, the definition of functional determinants as well as the normalisation of the Dirac delta function and the definition of functional derivatives. In many cases the $\ell$ dependence in these definitions will cancel out and the results of calculations using our conventions will reduce to those obtained using the standard definitions involving $g_{\mu\nu}$. Indeed, we see that the only dependence of $\ell$ in~\eqref{eq:cov eff act} that does not cancel is in the definition of the configuration space metric~\eqref{eq:config metric}.

However, the conventions we have laid out in this paper are essential if we want to keep both diffeomorphism invariance and frame invariance manifest. Without a frame invariant definition of the spacetime metric it would be impossible to define configuration space tensors that transform correctly. For example the standard functional derivative $\delta F/\delta\Phi^i$ does not transform as a configuration space vector, even if $F$ is a configuration space scalar. In addition, in order to obtain frame covariant correlation functions from the effective action we must use the covariant functional derivatives as defined in~\eqref{eq:config cov der}.

The freedom of choosing a preferred frame still exists in our formalism. However, it is now explicitly part of the content of the theory, captured by the model function~$\ell$, as opposed to being expressed by singling out a particular parametrisation. After all, the relation between the tensor field $g_{\mu\nu}$ and the metric of spacetime $\bar g_{\mu\nu}$ is a physical one and not just a convention. Two theories with the same classical action, but a different relation between $g_{\mu\nu}$ and $\bar g_{\mu\nu}$ cannot be related by a frame transformation \eqref{eq:grand trans} and will, in general, give rise to different quantum predictions.

Our formalism therefore draws a clear dividing line between the content of a theory and its representation. Once we have picked a particular form for the model functions $f,$ $k_{AB},$ $V,$ and~$\ell$, we have uniquely specified our QFT, and therefore all of its physical predictions. However, we may still change the representation of the theory by performing a frame transformation~\eqref{eq:grand trans}. The model functions will be different after this change of frame, but the QFT as defined by $\Gamma[{\bm \varphi}]$ will still have the same functional form as shown in~\eqref{eq:same gamma} and will make the same predictions. 

We note that when $\ell$ is treated as a model function, the definition~\eqref{eq:config metric} determines the configuration space metric in a way that does not depend on the parametrisation and does not rely on any `preferred' frame. Although we have not discussed it here in detail, previous works~\cite{DeWitt:1980jv,Vilkovisky:1984st,Rebhan:1986wp,Ellicott:1987ir,Fradkin:1983nw,Kunstatter:1986qa} have shown that any dependence on the gauge fixing condition can also be removed. Hence the configuration space metric for a scalar-tensor theory can be uniquely defined. Once the configuration space has been defined, the definition~\eqref{eq:cov eff act} fully determines the VDW effective action. Consequently, the VDW effective action is uniquely determined from the four model functions $f,$ $k_{AB},$ $V,$ and~$\ell$.

Since the covariant quantum effective action~\eqref{eq:cov eff act} is frame invariant and the ordinary effective action~\eqref{eq:standard effective action} is not, it is clear that they can only agree in at most one frame. This frame is one in which $\ell=1$ and additionally all the scalar fields are canonically normalised.
However, such a canonical frame does not exist for theories with intrinsic field-space curvature. Thus, for such theories, the usual approach is not suitable in any frame and we must adopt the formalism developed in this paper in order to maintain reparametrisation invariance.

This observation may be important for the development of a UV-complete quantum theory of gravity. The Einstein-Hilbert action for gravity~\eqref{eq:EH action} is non renormalizable, which has long prevented such a theory being constructed. As we have shown in Section~\ref{sec:GR}, General Relativity features a curved field space. This curvature alters the calculation of quantum corrections and must therefore be taken into account in order to UV-complete the theory.

By identifying the model function $\ell$, we have identified the source of the cosmological frame problem. It is not possible to write down a unique effective action without specifying the form of $\ell$ and any formalism that does not include this model function will have an inherent ambiguous choice of frame. Any frame transformation that does not take into account the transformation of $\ell$ will lead to a different theory with different quantum predictions. This implies that the classical action \emph{is not} sufficient to fully define a QFT.

In this paper, we have taken reparametrisation invariance as a fundamental guiding light. We argue that a theory should not depend on the way it is parametrised, and therefore Lagrangians related by a frame transformation are different expressions of the same underlying theory. Based on this idea, we have developed a formalism in which the invariance of physical predictions under such field reparametrisations is made manifest. Our formalism can be used to derive a quantum effective action that is manifestly invariant under frame transformations that include~$\ell$.

\setcounter{tocdepth}{-5}
\acknowledgments
The authors would like to thank Jack Holguin and Chris Shepherd for useful comments and discussion. We would also like to thank Daniel Martin for his help in writing the Mathematica code used to calculate covariant Feynman rules. KF is supported by the University of Manchester through the President's Doctoral Scholar Award. The work of AP and SK is supported by the Lancaster--Manchester--Sheffield Consortium for Fundamental Physics under STFC research grant ST/L000520/1.

\setcounter{tocdepth}{1}

\begin{appendices}
\renewcommand*{\thesubsection}{\alph{subsection}}
\renewcommand*{\thesubsubsection}{(\roman{subsubsection})}

\section{Theory with a Complex Scalar Field}
\label{sec:complex scalar}
   
As an example to highlight the parametrisation dependence of the standard formulation of QFTs, we consider the example of a single complex scalar field $\phi$ with action
\begin{equation}
S=\int d^4x\left[ \partial_\mu\phi\partial^\mu\phi- m^2\abs{\phi}^2-\lambda\abs{\phi}^4\right].\label{eq:complex action}
\end{equation}
We choose our parameters with $m^2<0$ so that the vacuum is 
\begin{equation}
\langle \phi\rangle=\frac{1}{\sqrt{2}}\rho_0\equiv\sqrt{\frac{-m^2}{2\lambda}}.\label{eq:complex vev}
\end{equation}

This theory has $U(1)$ symmetry $\phi\to e^{i\theta}\phi$, which is spontaneously broken by the vacuum~\eqref{eq:complex vev}. Therefore, the perturbations will have two modes, a massive Higgs mode and a massless Goldstone mode.

For simplicity we will assume a flat, static, background spacetime with Minkowski metric $\eta_{\mu\nu}=\mathrm{diag}\left(1,-1,-1,-1\right)$.

\subsection{Standard Approach: Linear Parametrisation}
\label{sec:standard lin}

The complex field $\phi$ contains two real degrees of freedom, which we can parametrise in terms of its real and imaginary parts as
\begin{equation}
\phi=\frac{\phi_1+i\phi_2}{\sqrt{2}}.\label{eq:linear param}
\end{equation}
In this parametrisation, the action~\eqref{eq:complex action} is
\begin{equation}
S=\int d^4x\bigg[\hf \partial_\mu\phi_1\partial^\mu\phi_1+\hf \partial_\mu\phi_2\partial^\mu\phi_2-\hf m^2(\phi_1^2+\phi_2^2)-\frac{\lambda}{4}(\phi_1^2+\phi_2^2)^2\bigg],\label{eq:linear action}
\end{equation}
and the vacuum~\eqref{eq:complex vev} is
\begin{align}
\langle \phi_1\rangle =\rho_0, \qquad
\langle \phi_2\rangle =0.
\end{align}
As stated before, the perturbations consist of a massive Higgs mode, corresponding to perturbations of $\phi_1$, and a massless Goldstone mode corresponding to perturbations of $\phi_2$.

\subsubsection{Effective Potential}

Let us start by calculating the one-loop correction to the effective action via~\eqref{eq:standard effective action1loop}. The inverse propagator for this theory is

\begin{equation}
\frac{\delta^2S}{\delta\phi^A(x)\delta\phi^B(y)}=\left(\begin{array}{cc}
-\partial^2-m^2-3\lambda\phi_1^2-\lambda\phi_2^2&-2\lambda\phi_1\phi_2\\
-2\lambda\phi_1\phi_2&-\partial^2-m^2-3\lambda \phi_2^2-\lambda\phi_1^2
\end{array}\right)\delta^{(4)}(x-y).
\end{equation}

Without loss of generality, we can use the $U(1)$ symmetry to set $\phi_2=0$. Thus, the one-loop effective action evaluated for a static configuration (the effective potential) in the \msbar renormalisation scheme is

\begin{align}\label{eq:V_eff linear}
\begin{aligned}
V_{\mathrm{eff}}(\varphi)  \equiv&-\frac{1}{V_4}\Gamma[\phi_1=\varphi,\phi_2=0] \\
=&V(\varphi)-\frac{i}{2}\ln\det G_{AB} \\
&+\frac{i}{2}\ln\det\left[\partial^2+m^2+3\lambda\varphi^2\right]+\frac{i}{2}\ln\det\left[\partial^2+m^2+\lambda\varphi^2\right] \\
=&\hf m^2\varphi^2+\frac{1}{4}\lambda\varphi^4+\frac{1}{64\pi^2}\bigg\{(m^2+3\lambda\varphi^2)^2\left[\ln\fb{m^2+3\lambda\varphi^2}{\mu^2}-\frac{3}{2}\right]\\
&\hphantom{\hf m^2\varphi^2+\frac{1}{4}\lambda\varphi^4+\frac{1}{64\pi^2}\bigg\{ }+(m^2+\lambda\varphi^2)^2\left[\ln\fb{m^2+\lambda\varphi^2}{\mu^2}-\frac{3}{2}\right]\bigg\},
\end{aligned}
\end{align}
where $V_4$ is the total four-volume of spacetime. Notice that, for a static configuration, $\ln\det G_{AB}=0$ in dimensional regularisation.

\subsubsection{Feynman Rules and Renormalisation}

The standard calculation~\eqref{eq:standard feyn rule} leads to the following Feynman rules:
\begin{align}
\label{eq:linear f rule 1}
\begin{aligned}
\SolidProp&=\frac{i}{p^2-m_1^2},&\DashedProp&=\frac{i}{p^2},\\
\SolidThreePoint&=-6i\lambda\rho_0,&\DashedThreePoint&=-2i\lambda\rho_0,\\
\SolidFourPoint&=-6i\lambda,&\DashedFourPoint&=-6i\lambda,&\HalfDashedFourPoint&=-2i\lambda,
\end{aligned}
\end{align}
where
\begin{equation}
m_1^2\equiv m^2+3\lambda\rho_0^2=-2m^2\label{eq:def m1}
\end{equation}
is the mass of the Higgs mode. Here we represent the Higgs mode~$\phi_1$ by a solid line and the Goldstone mode $\phi_2$ by a dashed line.

Let us use these Feynman rules to calculate the renormalisation of the Higgs mass. At one loop order we have
\begin{align}
\label{eq:linear higgs diagrams}
\begin{aligned}
i\Gamma_{\phi_1\phi_1}(p)=&\,\SolidProp+\SolidTadpole+\SolidDashedTadpole+\SolidTripleLoop\\
&+\DashedTripleLoop+\SolidLollipop+\SolidDashedLollipop\\
=&i(p^2-m_1^2)+\frac{3i\lambda}{(4\pi)^2} A(m_1^2)+\frac{i\lambda}{(4\pi)^2}A(0)+18i\frac{\lambda^2\rho_0^2}{(4\pi)^2}B_0(p^2,m_1,m_1)\\
&+2i\frac{\lambda^2\rho_0^2}{(4\pi)^2}B_0(p^2,0,0)-18i\frac{\lambda^2\rho_0^2}{(4\pi)^2m_1^2}A(m_1^2)-6i\frac{\lambda^2\rho_0^2}{(4\pi)^2m_1^2}A(0).
\end{aligned}
\end{align}
Here we have defined the following two integrals
\begin{align}
A(m^2)&\equiv \int\frac{d^4k}{i\pi^2}\frac{1}{k^2-m^2},\\
B_0(p^2,m_1,m_2)&\equiv \int \frac{d^4k}{i\pi^2} \frac{1}{k^2-m_1^2}\frac{1}{(p+k)^2-m_2^2},
\end{align}
which we can perform using dimensional regularisation scheme to give
\begin{align}
A(m^2)=&m^2\left[C_{\rm UV}+1-\ln\fb{m^2}{\mu^2}\right],\\
B_0(p^2,m_1,m_2)=&C_{\rm UV}-\int_0^1dx \ln\fb{m_1^2(1-x)+m_2^2 x-x(1-x)p^2}{\mu^2},
\end{align}
where $\mu$ is the renormalisation scale and
\begin{equation}
C_{\rm UV}=\frac{2}{4-D}-\gamma_{\rm E}+\ln(4\pi)
\end{equation}
is the UV divergence that is cancelled by counterterms in the \msbar renormalisation scheme. Here $D = 4 -2\epsilon$, and $\gamma_{\rm E}=0.577\ldots$ is the Euler--Mascheroni constant. We therefore have
\begin{align}
\begin{aligned}
\Gamma_{\phi_1\phi_1}(p)=&(p^2-m_1^2)+\frac{ \lambda m_1^2}{4\pi^2}\ln\fb{p^2}{\mu^2} \\
&-\frac{ \lambda m_1^2}{(4\pi)^2}\left[-4C_{\rm UV}+4+9\int_0^1 dx\ln\left(\frac{x(x-1)p^2+m_1^2}{\mu^2}\right)\right].
\end{aligned}\label{eq:linear higgs renorm}
\end{align}
Note that $A(0)=0$ and thus the third and final diagrams in~\eqref{eq:linear higgs diagrams} give no contribution.

From~\eqref{eq:linear higgs renorm} we see that there is no wavefunction renormalisation, as expected, and the beta function of the Higgs mass is
\begin{equation}
\beta_{m_1^2}=-\mu\frac{\partial { \widehat \Gamma}_2}{\partial\mu}=\frac{\lambda m_1^2}{2\pi^2}.\label{eq:linear beta m1}
\end{equation}

We can also calculate the Goldstone self energy. At one loop we have
\begin{align}
\begin{aligned}
i\Gamma_{\phi_2\phi_2}(p)=&\,\DashedProp+\DashedSolidTadpole+\DashedTadpole\\
&+\HalfDashedTripleLoop+\DashedSolidLollipop+\DashedLollipop\\
=&ip^2+\frac{i\lambda}{(4\pi)^2} A(m_1^2)+\frac{3i}{(4\pi)^2}\lambda A(0)\\
& +4i\frac{\lambda^2\rho_0^2}{(4\pi)^2}B_0(p^2,m_1,0)-6i\frac{\lambda^2\rho_0^2}{(4\pi)^2m_1^2}A(m_1^2)-2i\frac{\lambda^2\rho_0^2}{(4\pi)^2m_1^2}A(0)\\
=&ip^2-\frac{2i\lambda m_1^2}{16\pi^2}\int_0^1dx\ln\left(1-\frac{xp^2}{m^2_1}\right).
\end{aligned}\label{eq:linear goldstone renorm}
\end{align}
Since~\eqref{eq:linear goldstone renorm} has no dependence on $\mu$, the Goldstone mass is not renormalised and remains zero in accordance with Goldstone's theorem.

Finally, let us compute the coupling renormalisation using the Callan Symanzic equation~\cite{Callan:1970yg,Symanzik:1970rt}
\begin{equation}
\left[\mu\frac{\partial}{\partial\mu}+\beta_\lambda\frac{\partial}{\partial\lambda}+\beta_{m^2}\frac{\partial}{\partial m^2}\right]\widetilde{V}_{\rm eff}=0.\label{eq:callan symanzic}
\end{equation}
where
\begin{equation}
\widetilde{V}_{\rm eff}(\varphi)=V_{\mathrm{eff}}(\varphi)-V_{\mathrm{eff}}(0)
\end{equation}
is the modified effective potential. From the expression for $V_{\mathrm{eff}}$ in~\eqref{eq:V_eff linear} we have, at leading order,
\begin{equation}
\frac{1}{4}\beta_\lambda\varphi^4-\frac{1}{4}\beta_{m_1^2}\varphi^2 -\frac{(m^2+3\lambda\varphi^2)^2+(m^2+\lambda\varphi^2)^2-2m^4}{32\pi^2}=0,
\label{eq:linear CS}
\end{equation}
where we have used the identity $\beta_{m_1^2}=-2\beta_{m^2}$, which derives from~\eqref{eq:def m1}.

Rearranging, and using the expression for $\beta_{m_1^2}$ from~\eqref{eq:linear beta m1} we see that the beta function for the coupling renormalisation, evaluated at the vacuum~$\varphi=\rho_0$, is
\begin{equation}
\beta_\lambda=\frac{5}{4\pi^2}\lambda^2.\label{eq:linear beta lambda}
\end{equation}

\subsection{Standard Approach: Non-Linear Parametrisation}
\label{sec:standard nlin}

Alternatively we could have used a non-linear parametrisation of the complex field
\begin{equation}
\phi=\frac{1}{\sqrt{2}}\rho e^{i\frac{\sigma}{\rho_0}}.\label{eq:non linear param}
\end{equation}
In this parametrisation the action~\eqref{eq:complex action} is
\begin{equation}
S=\int d^4x\left[\hf \partial_\mu\rho\partial^\mu\rho+\hf\fb{\rho}{\rho_0}^2 \partial_\mu\sigma\partial^\mu\sigma-\hf m^2\rho^2-\frac{\lambda}{4}\rho^4\right]\label{eq:non linear action}
\end{equation}
and the vacuum~\eqref{eq:complex vev} is
\begin{align}
\langle \rho\rangle =\rho_0,  \qquad
\langle \sigma\rangle =0.
\end{align}
In this parametrisation the Higgs mode is in the direction of $\rho$ and the Goldstone mode is in the direction of $\sigma$.

\subsubsection{Effective Potential}

Let us calculate the one-loop effective action in this parametrisation using~\eqref{eq:standard effective action1loop}. The inverse propagator in this parametrisation is
\begin{equation}
\frac{\delta^2S}{\delta\phi^a(x)\delta\phi^b(y)}=\left(\begin{array}{cc}
-\partial^2+\partial_\mu\sigma\partial^\mu\sigma-m^2-3\lambda\rho^2&-2\partial_\mu\rho\partial^\mu\sigma-2\rho\partial^2\sigma-2\rho\partial_\mu\sigma\partial^\mu\\
-2\partial_\mu\rho\partial^\mu\sigma-2\rho\partial^2\sigma-2\rho\partial_\mu\sigma\partial^\mu&-2\rho\partial_\mu\rho\partial^\mu-\rho^2\partial^2
\end{array}\right)\delta^{(4)}(x-y).
\end{equation}

As before we can, without loss of generality, use the $U(1)$ symmetry to set $\sigma=0$. Again, we will consider a static configuration in order to calculate the effective action. In the \msbar scheme, this is given by
\begin{align}
\begin{aligned}
V_{\mathrm{eff}}(\varphi)\equiv &-\frac{1}{V_4}\Gamma[\rho=\varphi,\sigma=0] \\
&=\hf m^2\varphi^2+\frac{\lambda}{4}\varphi^4-\ln\det\left[\partial^2+m^2+3\lambda\varphi^2\right]-\ln\det\left[\varphi^2\partial^2\right]\\
&=\hf m^2\varphi^2+\frac{\lambda}{4}\varphi^4+\frac{1}{64\pi^2}\left\{(m^2+3\lambda\varphi^2)^2\left[\ln\fb{m^2+3\lambda\varphi^2}{\mu^2}-\frac{3}{2}\right]\right\}. 
\end{aligned}
\label{eq:V_eff non-linear}
\end{align}

Notice that~\eqref{eq:V_eff linear} and~\eqref{eq:V_eff non-linear} differ off shell, highlighting the parametrisation dependence of the standard effective action. However, on shell when $\varphi=\rho_0=\sqrt{-m^2/\lambda}$, the two expressions agree.

Note that if we had instead taken $m^2>0$, the vacuum would lie at $\varphi=0$. Surprisingly, in this case~\eqref{eq:V_eff linear} and~\eqref{eq:V_eff non-linear} do not agree even on shell. This is due to a peculiarity with the particular coordinate chart~\eqref{eq:non linear param}. The point $\phi=0$ is multiply covered by this chart and therefore represents a coordinate singularity -- a point where the chart cannot be trusted.

To rectify this problem we can define an offset parametrisation $\phi=\frac{1}{\sqrt{2}}\left(\tilde{\rho} e^{i\tilde{\sigma}} -\delta\right)$ so that the vacuum $\phi=0$ is no longer at the singular point. In the offset parametrisation, the effective potential is
\begin{align}
\begin{aligned}
V_{\mathrm{eff}}(\tilde{\rho}={\widetilde \varphi},\tilde{\sigma}=0) =&\hf m^2({\widetilde \varphi}-\delta)^2+\frac{\lambda}{4}({\widetilde \varphi}-\delta)^4\\
&+\frac{(m^2+3\lambda({\widetilde \varphi}-\delta)^2)^2}{64\pi^2}\left[\ln\fb{m^2+3\lambda({\widetilde \varphi}-\delta)^2}{\mu^2}-\frac{3}{2}\right]\\
&+\frac{\delta^2}{{\widetilde \varphi}^2}\frac{(m^2+\lambda({\widetilde \varphi}-\delta)^2)^2}{64\pi^2}\left[\ln\left(\frac{\delta}{{\widetilde \varphi}}\frac{ m^2+\lambda({\widetilde \varphi}-\delta)^2}{\mu^2}\right)-\frac{3}{2}\right], 
\end{aligned}\label{eq:Veff offset}
\end{align}
which we can see \emph{does} agree with~\eqref{eq:V_eff linear} at ${\widetilde \varphi}=\delta$. Thus, in order to calculate the effective action~\eqref{eq:V_eff non-linear} at $\varphi=0$, we should take the limit ${\widetilde \varphi}\to\delta \to 0$ in~\eqref{eq:Veff offset}, which gives us
\begin{equation}
V_{\mathrm{eff}}(0)=\frac{2m^2}{64\pi^2}\left[\ln\fb{ m^2}{\mu^2}-\frac{3}{2}\right]
\end{equation}
in agreement with~\eqref{eq:V_eff linear}. This expression will be needed for the Callan--Symanzic equation.

\subsubsection{Feynman Rules and Renormalisation}

The standard Feynman rules from~\eqref{eq:standard feyn rule} in this parametrisation are:
\begin{align}
\begin{aligned}
\SolidProp&=\frac{i}{p^2-m_1^2},&\DashedProp&=\frac{i}{p^2},\\
\SolidThreePoint&=-6i\lambda\rho_0,&\DashedThreePointKs&=-\frac{2i}{\rho_0}k_1\cdot k_2, \\
\SolidFourPoint&=-6i\lambda,&\HalfDashedFourPointHalfKs&=-\frac{2i}{\rho_0^2}k_1\cdot k_2.
\end{aligned}
\end{align}
As before, the solid line represents the Higgs mode and the dashed line represents the Goldstone mode. As expected these Feynman rules are different from~\eqref{eq:linear f rule 1}, showing explicitly the parametrisation non-invariance of this approach.

If we calculate the Higgs mass renormalisation with this parametrisation, we will find
\begin{align}
\begin{aligned}
i\Gamma_{\rho\rho}(p)=&\,\SolidProp+\SolidTadpole+\SolidDashedTadpole+\SolidTripleLoop\\
&+\DashedTripleLoop+\SolidLollipop+\SolidDashedLollipop\\
= &\,i(p^2-m_1^2)+\frac{3i\lambda}{(4\pi)^2} A(m_1^2)+\frac{2i}{\rho_0}\int\frac{d^4k}{(2\pi)^4}+18i\frac{\lambda^2\rho_0^2}{(4\pi)^2}B_0(p^2,m_1,m_1)\\
&+\frac{i}{2}\frac{p^4}{(4\pi)^2\rho_0^2}B_0(p^2,0,0)-18i\frac{\lambda^2\rho_0^2}{(4\pi)^2 m_1^2}A(m_1^2)+6i\frac{\lambda^2\rho_0^2}{m_1^2}\int\frac{d^4k}{(2\pi)^4}\\
=&\,i(p^2-m_1^2)+\frac{3\lambda m_1^2+\lambda\frac{p^4}{m_1^4}}{(4\pi)^2}\ln\fb{p^2}{\mu^2}\\
&+\frac{i\lambda m_1^2}{(4\pi)^2}\bigg[\left(3+\frac{p^4}{m_1^4}\right)C_{\rm UV}-6+2\frac{p^4}{m_1^4}-9\int_0^1dx\ln\left(\frac{x(x-1)p^2+m_1^2}{\mu^2}\right)\bigg].
\end{aligned}\label{eq:non linear higgs renorm}
\end{align}
We see that, as expected, this differs from~\eqref{eq:linear higgs renorm} off shell. In fact, due to the presence of the $p^4$ divergence, this theory is naively non-renormalisable. However, if we only consider on-shell momentum, so that $p^2=m_1^2$ the two expressions,~\eqref{eq:linear higgs renorm} and~\eqref{eq:non linear higgs renorm} are equal. As a result, the beta function will be given by~\eqref{eq:linear beta m1}.

We can also calculate the Goldstone mass renormalisation
\begin{align}
\begin{aligned}
i\Gamma_{\sigma\sigma}(p)&=\,\DashedProp+\DashedSolidTadpole+\DashedSolidLollipop+\DashedLollipop+\HalfDashedTripleLoop\\
=  &ip^2-\frac{i p^2}{(4\pi)^2\rho_0^2}A(m_1^2)+6i\frac{\lambda p^2}{(4\pi)^2m_1^2} A(m_1^2)\\
&-2i\frac{p^2}{\rho_0^2}\int\frac{d^4k}{(2\pi)^4}+\left[i\frac{3p^2-m_1^2}{(4\pi)^2\rho_0^2}A(m_1^2)+i\frac{(p^2-m_1^2)^2}{(4\pi)^2\rho_0^2}B_0(p^2,m_1,0)\right]\\
=&ip^2+i\frac{ 5p^2-m_1^2}{16\pi^2\rho_0^2}m_1^2\left[C_{\rm UV}+1-\ln\fb{m^2}{\mu^2}\right]\\
&+i\frac{(p^2-m_1^2)^2}{16\pi^2\rho_0^2}\bigg[C_{\rm UV}-\int_0^1dx\ln\left(\frac{(1-x)m_1^2-x(1-x)p^2}{\mu^2}\right)  \bigg]. 
\end{aligned}
\end{align}
As before, this expression differs from the expression obtained using the linear parametrisation~\eqref{eq:linear goldstone renorm} and also contains non-renormalisable terms. However, on-shell, when $p^2=0$, we have
\begin{equation}
\Gamma_{\sigma\sigma}(p^2=0)=0
\end{equation}
in agreement with~\eqref{eq:linear goldstone renorm} and the Goldstone mass is not renormalised as expected by Goldstone's theorem.

Finally we look at the coupling renormalisation, which we shall calculate through the Callan-Symanzic equation~\eqref{eq:callan symanzic} as before. Using the expression~\eqref{eq:V_eff non-linear} for the effective action gives us
\begin{equation}
-\frac{(m^2+3\lambda\varphi^2)^2-2m^4}{32\pi^2}+\frac{1}{4}\beta_\lambda\varphi^4-\frac{1}{4}\beta_{m_1^2}\varphi^2=0.
\end{equation}
On shell when $\varphi=\rho_0$, this becomes identical to~\eqref{eq:linear CS}, which means that the beta function for the coupling renormalisation is
\begin{equation}
\beta_\lambda=\frac{5}{4\pi^2}\lambda^2
\end{equation}
as before.\footnote{Note that we had to use~\eqref{eq:Veff offset} to calculate $V_{\mathrm{eff}}(\varphi=0)$. Had we used~\eqref{eq:V_eff non-linear} instead, the two results would not have agreed. As stated earlier, this is because the parametrisation~\eqref{eq:non linear param} features a coordinate singularity at $\rho=0$ and so cannot be trusted there.}

Although the two approaches led to several differences in the intermediate, off-shell results, as the above calculation demonstrates all physical observables are the same regardless of the parametrisation.  

\vspace{1.3em}

\subsection{Covariant Approach}

We have shown in the main text how to alleviate the parametrisation dependence of quantum calculations by using an explicitly covariant formalism. Let us now repeat the above calculations using this formalism to show how parametrisation invariance is maintained.

For the linear parametrisation~\eqref{eq:linear param}, the field-space is trivial, and so there is no difference between the covariant approach and the standard (ordinary) approach. Thus, the VDW effective potential will be~\eqref{eq:V_eff linear}, the covariant Feynman rules will give~\eqref{eq:linear f rule 1}, and the renormalisation group calculations will be identical to those in Section~\ref{sec:standard lin}.

We will therefore focus on the non-linear parametrisation~\eqref{eq:non linear param}. In this parametrisation, the configuration-space metric~\eqref{eq:scalar config metric} is
\begin{equation}
\G_{ab}=\begin{pmatrix}
1&0\\
0& (\rho/\rho_0)^2
\end{pmatrix}\delta^{(4)}(x_a-x_b)
\end{equation}
and the non-zero configuration-space Christoffel symbols can be calculated as
\begin{align}
\Gamma^{\rho(z)}_{\sigma(x)\sigma(y)}&=-\frac{\rho(z)}{\rho_0^2}\delta^{(4)}(z-x)\delta^{(4)} (z-y),\\
\Gamma^{\sigma(z)}_{\rho(x)\sigma(y)}&=\frac{1}{\rho(z)}\delta^{(4)} (z-x)\delta^{(4)} (z-y).
\end{align}

\subsubsection{Vilkovisky DeWitt Effective Potential}

Let us first calculate the Vilkovisky DeWitt effective action for this theory using~\eqref{eq:vdw 1 loop}. The covariant $2\times 2$ inverse propagator is
\begin{equation}\label{eq:modified prop}
\nabla_a\nabla_b S=\left(\begin{array}{cc}
-\partial^2+\partial_\mu\sigma\partial^\mu\sigma-m^2-3\lambda\rho^2&\rho\partial_\mu\sigma\partial^\mu\\
\rho\partial_\mu\sigma\partial^\mu & -\rho\partial_\mu\rho\partial^\mu-\rho^2\partial^2+\rho^2\partial_\mu\sigma\partial^\mu\sigma -m^2\rho^2-\lambda\rho^4
\end{array}\right)\delta(x_I-x_J),
\end{equation}
where $\phi^a = (\rho,\sigma)$. 
As before, we can use the $U(1)$ symmetry to set $\sigma=0$ without loss of generality. We will also consider a static configuration as before. Therefore, the one-loop VDW effective potential in the \msbar scheme reads
\begin{align}
\begin{aligned}
V_{\mathrm{eff}}(\varphi)\equiv &-\frac{1}{V_4}\Gamma[\rho=\varphi,\sigma=0] \label{eq:V_VdW} \\
=&\hf m^2\varphi^2+\frac{\lambda}{4}\varphi^4-\ln\det\left[\partial^2+m^2+3\lambda\varphi^2\right]\\
&-\ln\det\left[\partial^2+m^2+\lambda\varphi^2\right]-\ln(\varphi^2)+\ln\det[G_{ab}] \\
=&\hf m^2\varphi^2+\frac{1}{4}\lambda\varphi^4+\frac{1}{64\pi^2}\bigg\{(m^2+3\lambda\varphi^2)^2\left[\ln\fb{m^2+3\lambda\varphi^2}{\mu^2}-\frac{3}{2}\right]\\
&\hphantom{\hf m^2\varphi^2+\frac{1}{4}\lambda\varphi^4+\frac{1}{64\pi^2}\bigg\{}+(m^2+\lambda\varphi^2)^2\left[\ln\fb{m^2+\lambda\varphi^2}{\mu^2}-\frac{3}{2}\right]\bigg\},
\end{aligned}
\end{align}
where $V_4$ is the four-volume. Observe that the expression \eqref{eq:V_VdW} is identical to \eqref{eq:V_eff linear}. As expected, the Vilkovisky DeWitt effective action is independent of parametrisation.

\subsubsection{Covariant Feynman Rules and Renormalisation}

With the help of~\eqref{eq:cov feyn rule}, we can calculate the covariant Feynman rules for this theory. We find them to be
\begin{align}
\begin{aligned}
\SolidProp&=\frac{i}{p^2-m_1^2},&\DashedProp&=\frac{i}{p^2},\\
\SolidThreePoint&=-6i\lambda\rho_0,&\DashedThreePoint&=-2i\lambda\rho_0,\\
\SolidFourPoint&=-6i\lambda,&\DashedFourPoint&=-6i\lambda, &\HalfDashedFourPoint&=-2i\lambda.
\end{aligned}
\end{align}
By construction, these Feynman rules are identical to~\eqref{eq:linear f rule 1} and thus all RG calculation are identical both on and off shell.

\section{Curved Field-Space Example}
\label{sec:curved field}

We wish to consider a simple toy model with genuine field-space curvature in order to study the effect this has on the quantum observables. Since it is impossible to have curvature in one dimension, we consider a theory with two fields $\rho$ and $\sigma$, and take $\sigma$ to be an angular variable with a shift symmetry. In order to avoid ghosts, the metric of the field space must be positive-definite. Consequently, we take our field-space metric to be
\begin{equation}
G_{AB}=\left(\begin{array}{cc} 1&0\\0& (\rho/\rho_0)^{2n}
\end{array}\right).\label{eq:general n metric}
\end{equation}

The non-zero Christoffel symbols of this metric are easily calculated 
\begin{align}
\Gamma^\rho_{\sigma\sigma} =-n\frac{\rho^{2n-1}}{\rho_0^{2n}}, 
\qquad
\Gamma^\sigma_{\rho\sigma} =\frac{n}{\rho} \, .
\end{align}
From these, we obtain the non-zero components of the field-space Riemann tensor
\begin{align}
\begin{aligned}
R^\rho_{\sigma\rho\sigma}&=  -\frac{n(n-1)}{\rho^2}\frac{\rho^{2n}}{\rho_0^{2n}}, 
  & 
R^\rho_{\sigma\sigma\rho}&= \frac{n(n-1)}{\rho^2}\frac{\rho^{2n}}{\rho_0^{2n}},\\
R^\sigma_{\rho\rho\sigma}&=\frac{n(n-1)}{\rho^2},
  &
R^\sigma_{\rho\sigma\rho}&=-\frac{n(n-1)}{\rho^2}.
\end{aligned}
\end{align}
We see that provided $n\neq0,1$, the Riemann tensor is non-zero and thus the field space is curved. 
Notice that $n=0$ and $n=1$ correspond to the two flat field space examples we have looked at already in Appendix~\ref{sec:standard lin} and~\ref{sec:standard nlin} respectively.

The simplest model with curvature is therefore the case $n=2$, which has the Lagrangian
\begin{equation}
\L=\hf \partial_\mu\rho\partial^\mu\rho+\frac{1}{2} \fb{\rho}{\rho_0}^{4}\partial_\mu\sigma\partial^\mu\sigma-\hf m^2 \rho^2-\frac{\lambda}{4}\rho^4.\label{eq:RG 4 action}
\end{equation}
We will consider the symmetry-broken vacuum
\begin{equation}
\langle\rho\rangle=\rho_0\equiv\sqrt{\frac{-m^2}{\lambda}}, \label{eq:RG curved rho0}
\end{equation}
as we have done before.

\subsection{Standard Approach}
\label{sec:standard curved}

We start by calculating the renormalisation group flow in the standard way by looking at the standard Feynman rules~\eqref{eq:standard feyn rule}. For the Lagrangian~\eqref{eq:RG 4 action}, these are given by
\begin{align}
\SolidProp&=\frac{i}{p^2-m_1^2},&\DashedProp&=\frac{i}{p^2},\\
\SolidThreePoint&=-6i\lambda\rho_0,&\DashedThreePointKs&=-4i\frac{k_1\cdot k_2}{\rho_0},\nonumber \\
\SolidFourPoint&=-6i\lambda,&\HalfDashedFourPointKs&=-12i\frac{k_3\cdot k_4}{\rho_0^2}, \nonumber
\end{align}
where $m_1^2\equiv m^2+3\lambda\rho_0^2$ as before. Higher order interactions also exist, however they will not be necessary for the one-loop calculations we will perform.

First, we will calculate the self-energy of the $\rho$ field. This is given schematically by
\begin{align}
\begin{aligned}
i\Gamma_{\rho\rho}(p)=&\SolidProp+\SolidTadpole+\SolidLollipop +\SolidDashedLollipop\\
&+\SolidTripleLoop +\SolidDashedTadpole+\DashedTripleLoop
\end{aligned}
\end{align}
and calculated to be 
\begin{align}
\begin{aligned}
i\Gamma_{\rho\rho}(p)=&i(p^2-m_1^2)+\frac{3i\lambda}{(4\pi)^2} A(m_1^2)-18i\frac{\lambda^2\rho_0^2}{(4\pi)^2m_1^2}A(m_1^2)+12\frac{\lambda}{m_1^2}\int\frac{d^4k}{(2\pi)^4}\\
&+18i\frac{\lambda^2\rho_0^2}{(4\pi)^2}B_0(p^2,m_1,m_1)-6\frac{1}{\rho_0^2}\int\frac{d^4k}{(2\pi)^4} +\frac{2ip^4}{(4\pi)^2\rho_0^2}B_0(p^2,0,0) \\
=&i(p^2-m_1^2)+\frac{i\lambda m_1^2}{(4\pi)^2}\bigg[ \left(3+4\frac{p^4}{m_1^4}\right)C_{\rm UV} +6\ln\fb{m_1^2}{\mu^2}-4\frac{p^4}{m_1^4}\ln\fb{p^2}{\mu^2}\\
&\hphantom{i(p^2-m_1^2)+\frac{i\lambda m_1^2}{(4\pi)^2}\bigg[}-9\int_0^1dx\ln\fb{m_1^2-x(1-x)p^2}{\mu^2}-6+8\frac{p^4}{m_1^4}\bigg] . 
\end{aligned}
\end{align}
As expected, the non-renormalisability of the theory leads to a UV-divergent term proportional to $p^4$, which cannot be absorbed into a counterterm. In order to compare to the covariant approach, we calculate the on-shell self energy as follows:
\begin{equation}
\Gamma_{\rho\rho}(p^2\!=\!m_1^2)=\!\frac{ \lambda m_1^2}{(4\pi)^2}\!\left[-7C_{\rm UV}-\!7\ln\!\fb{m_1^2}{\mu^2}\!+20\!-\!3\sqrt{3}\pi\right]\!.\label{eq:curved standard higgs mass on shell}
\end{equation}

We now compare to the Goldstone self-energy, which is given schematically by
\begin{align}
i\Gamma_{\sigma\sigma}(p)=&\,\DashedProp+\DashedSolidTadpole+\DashedSolidLollipop+\DashedLollipop+\HalfDashedTripleLoop
\end{align}
and calculated to be
\begin{align}
\begin{aligned}
i\Gamma_{\sigma\sigma}(p)=&ip^2+\frac{6ip^2}{(4\pi)^2\rho_0^2}A(m_1^2)-12i\frac{p^2\lambda}{(4\pi)^2m_1^2}A(m_1^2)-16\frac{p^2}{\rho_0^2m_1^2}\int\frac{d^4k}{(2\pi)^4}\\
&+4i\bigg[\frac{3p^2-m_1^2}{(4\pi)^2\rho_0^2}A(m_1^2)+\frac{(p^2-m_1^2)^2}{(4\pi)^2\rho_0^2}B_0(p^4,m_1,0)\bigg]\\
=&ip^2+4i\frac{ 3p^2-m_1^2}{(4\pi)^2\rho_0^2}m_1^2\left[C_{\rm UV}+1-\ln\fb{m_1^2}{\mu^2}\right]\\
&+4i\frac{(p^2-m_1^2)^2}{(4\pi)^2\rho_0^2}\bigg[C_{\rm UV}+1-\int_0^1dx\ln\fb{xp^2-m_1^2}{\mu^2}\bigg]. 
\end{aligned}
\end{align}
As before, due to the non-renormalisability of the theory, this expression contains divergences that cannot be absorbed by a counterterm. However, on-shell we have $p^2=0$ and the expression reduces to
\begin{equation}
\Gamma_{\sigma\sigma}(p^2=0)=0,\label{eq:curved standard goldstone mass on shell}
\end{equation}
implying that the Goldstone boson receives no correction to its mass as expected.

Finally, it is instructive to calculate the tree-level S-matrix element for $\rho\rho\to\sigma\sigma$. Taking into account the contributing diagrams, find
\begin{align}
\begin{aligned}
i \mathcal{M}(\rho\rho \to \sigma \sigma)  &= \HalfDashedSMatrix\\
&=\HalfDashedFourPoint+\HalfDashedSTree+\HalfDashedTTree+\HalfDashedUTree \\
&=-6i\frac{s}{\rho_0^2}-6i\frac{sm_1^2}{\rho_0^2(s-m_1^2)}-4i\frac{(m_1^2-t)^2}{\rho_0^2t}-4i\frac{(m_1^2-u)^2}{\rho_0^2u}\\
&=-\frac{2i}{\rho_0^2}\left[3\frac{s^2}{s-m_1^2}+2\frac{(m_1^2-t)^2}{t}+2\frac{(m_1^2-u)^2}{u}\right], 
\end{aligned}\label{eq:higgs goldstone standard}
\end{align}
where $s=(k_1 + k_2)^2, t = (k_1 - k_3)^2, u = (k_1 - k_4)^2$ are the standard Mandelstam variables. Note that in the high energy limit, $\mathcal{M}(\rho\rho \to \sigma \sigma)  \propto -2s/\rho_0^2$.

\subsection{Covariant Approach}

We now perform analogous calculations in the covariant formalism using~\eqref{eq:cov feyn rule}. Here we show a limited set of the covariant Feynman rules for this theory
\begin{align}
\begin{aligned}
\SolidProp&=\frac{i}{p^2-m_1^2},   &\DashedProp&=\frac{i}{p^2},\\
\SolidThreePoint&=-6i\lambda\rho_0,  &\DashedThreePoint&=-4i\lambda\rho_0,\\
\SolidFourPoint&=-6i\lambda,  &\DashedFourPoint&=-24i\lambda.
\end{aligned}
\end{align}
There are also an infinite set of higher order vertices, which we do not calculate since they do not affect the one-loop calculations we make in this section.

Finally, there is the $\rho\rho\sigma\sigma$ vertex. Due to the curvature of the field space there is an ambiguity in the order in which the covariant derivatives are taken when calculating this vertex. In Section~\ref{sec:cov feyn rules}, we argued that the correct approach was to symmetrise over all possible orderings. Nevertheless, we calculate each ordering explicitly. We have
\begin{align}
\begin{aligned}
\underset{\rho\rho\sigma\sigma}{\HalfDashedFourPointKs}\!\!\!&\equiv \nabla_\rho\nabla_\rho\nabla_\sigma\nabla_\sigma S=-4i\lambda-4i\frac{k_1\cdot k_2}{\rho_0^2},\\
\underset{\sigma\sigma\rho\rho}{\HalfDashedFourPointKs}\!\!\!&\equiv \nabla_\sigma\nabla_\sigma\nabla_\rho\nabla_\rho S=4i\lambda-4i\frac{k_3\cdot k_4}{\rho_0^2},\\
\underset{\rho\sigma\rho\sigma,\;\rho\sigma\sigma\rho}{\HalfDashedFourPointKs}\!\!\!&\equiv \nabla_\rho\nabla_\sigma\nabla_\rho\nabla_\sigma S=2i\frac{k_1\cdot \left(k_3+k_4-k_2\right)}{\rho_0^2},\\
\underset{\sigma\rho\rho\sigma,\;\sigma\rho\sigma\rho}{\HalfDashedFourPointKs}\!\!\!&\equiv \nabla_\sigma\nabla_\rho\nabla_\rho\nabla_\sigma S=4i\lambda+2i\frac{k_3\cdot \left(k_1+k_2-k_4\right)}{\rho_0^2},
\end{aligned}
\end{align}
where the ordering is denoted under the diagram. In the above, we did not display the other orderings of the two individual $\rho$ and $\sigma$ particles when calculating these rules, which may be obtained by exchanging $k_1\leftrightarrow k_2$ and/or $k_3\leftrightarrow k_4$. 

It is interesting to calculate the form of the vertices when taking all external particles to be on-shell. We set
\begin{align}
k_1^2=k_2^2 =m_1^2=2\lambda\rho_0^2, \qquad
k_3^2=k_4^2 =0.
\end{align}
Conservation of momentum then implies that
\begin{align}
0 &=k_1+k_2+k_3+k_4, &
k_1\cdot k_2&=k_3\cdot k_4-m_1^2,\nonumber\\
k_1\cdot k_3&=k_2\cdot k_4, &
k_1\cdot k_4&=k_2\cdot k_3.
\end{align}
Employing these relations, we see that on-shell, all six orderings are equal
\begin{equation}
\label{sixorderings}
\underset{\rho\rho\sigma\sigma}{\HalfDashedFourPointKs}=\underset{\sigma\sigma\rho\rho}{\HalfDashedFourPointKs}=\underset{\rho\sigma\rho\sigma,\;\rho\sigma\sigma\rho}{\HalfDashedFourPointKs}=\underset{\sigma\rho\rho\sigma,\;\sigma\rho\sigma\rho}{\HalfDashedFourPointKs}=4i\lambda-4i\frac{k_3\cdot k_4}{\rho_0^2}.
\end{equation}
Notice that the expression on \eqref{sixorderings} is invariant under $k_1\leftrightarrow k_2$ and $k_3\leftrightarrow k_4$.

We have found that ordering does not matter when the particles are on shell. However, any quantum calculation will involve off-shell particles and for these, the ordering will make a difference. It is therefore important to use the fully symmetrised rule
\begin{equation}
\HalfDashedFourPointKs =\frac{2i\lambda}{3}-\frac{2i}{3}\frac{(k_1-k_3)\cdot(k_2-k_4)+(k_1-k_4)\cdot(k_2-k_3)}{\rho_0^2}, \label{eq:average feyn rule}
\end{equation}
as discussed in Section~\ref{sec:cov feyn rules}.

Let us now calculate the $\rho\rho$ and $\sigma\sigma$ self-energy as we did above. For the $\sigma\sigma$ self-energy, we obtain
\begin{align}
\begin{aligned}
i\Gamma_{\rho\rho}(p)=&\,\SolidProp+\SolidTadpole+\SolidLollipop+\SolidDashedLollipop\\
&+\SolidTripleLoop+\DashedTripleLoop+\SolidDashedTadpole\\
=&i(p^2-m_1^2)+\frac{3i\lambda}{(4\pi)^2} A(m_1^2)-18i\frac{\lambda^2\rho_0^2}{(4\pi)^2m_1^2}A(m_1^2)\\
&-12i\frac{\lambda^2\rho_0^2}{(4\pi)^2m_1^2}A(0)+18i\frac{\lambda^2\rho_0^2}{(4\pi)^2}B_0(p^2,m_1,m_1)\\
&+8i\frac{\lambda^2\rho_0^2}{(4\pi)^2}B_0(p^2,0,0)+\frac{i}{2}\left[\frac{1}{(4\pi)^2}\left(\frac{2}{3}\lambda+\frac{p^2}{\rho_0^2}\right)A(0)+\frac{1}{\rho_0^2}\int \frac{d^4k}{(2\pi)^4}\right]\\
=&i(p^2-m_1^2)+\frac{i\lambda m_1^2}{(4\pi)^2}\left[7C_{\rm UV}+ 6\ln\fb{m_1^2}{\mu^2}+4\ln\fb{p^2}{\mu^2}\right.\\
&\left.-2+9\int_0^1 dx\ln\fb{m_1^2 - x(1-x)p^2+ }{\mu^2}\right].
\end{aligned}\label{eq:curved cov higgs diagrams}
\end{align}
Notice that in the covariant approach, there is no non-renormalisable divergence.  If we set the particle on-shell, we get
\begin{equation}
\Gamma_{\rho\rho}(p^2=m_1^2)=\frac{ \lambda m_1^2}{(4\pi)^2}\left[7C_{\rm UV}-7\ln\fb{m_1^2}{\mu^2}+20-3\sqrt{3}\pi\right].
\end{equation}
This is in agreement with~\eqref{eq:curved standard higgs mass on shell}.

In the previous calculation, only the final diagram of~\eqref{eq:curved cov higgs diagrams} depends on the ordering of the Feynman rule and it vanishes regardless of the ordering. As such, the ordering was not really tested in this calculation. Let us instead calculate the self-energy of the $\sigma$ field, which will test the ordering. This is given by
\begin{align}
\begin{aligned}
i\Gamma_{\sigma\sigma}(p)
 =&\DashedDotProp \\
=&\DashedProp        + \DashedTadpole +\DashedSolidLollipop + \DashedLollipop\\
&+  \HalfDashedTripleLoop+ \DashedSolidTadpole .
\end{aligned} 
\end{align}

Before completing this calculation, let us focus on the last diagram, which is the only one in which the ordering makes a difference. Although we will eventually symmetrise over all possible orderings, let us first consider them all individually:
\begin{align}
\begin{aligned}
\underset{\rho\rho\sigma\sigma,\;\rho\sigma\rho\sigma,\;\rho\sigma\sigma\rho}{\DashedSolidTadpole}&=-\frac{2i\lambda}{(4\pi)^2} A(m_1^2),\\
\underset{\sigma\sigma\rho\rho}{\DashedSolidTadpole}&=-\frac{2i\lambda}{(4\pi)^2}\left(1+\hf\frac{p^2}{m_1^2}\right)A(m_1^2),\\
\underset{\sigma\rho\rho\sigma,\;\sigma\rho\sigma\rho}{\DashedSolidTadpole}&=-\frac{2i\lambda}{(4\pi)^2}\left(1+\frac{1}{4}\frac{p^2}{m_1^2}\right) A(m_1^2).
\end{aligned}
\end{align}
Note that the different orderings of the above diagrams lead to different results. However, these results converge to a single expression when the external particles are taken to be on-shell, i.e $p^2=0$, despite the fact that the particle in the loop is off-shell. For the diagram with off-shell external particles, we will use the symmetrised Feynman rule, which gives
\begin{equation}
\DashedSolidTadpole=-\frac{2i}{(4\pi)^2}\lambda\left(1+\frac{1}{6}\frac{p^2}{m_1^2}\right) A(m_1^2).
\end{equation}

In this way, we find
\begin{align}
\begin{aligned}
 \Gamma_{\sigma\sigma}=&  p^2+\frac{12 \lambda}{(4\pi)^2} A(0)-12 \frac{\lambda^2\rho_0^2}{(4\pi)^2m_1^2} A(m_1^2) -8 \frac{\lambda^2\rho_0^2}{(4\pi)^2m_1^2}A(0)\\
 &+16 \frac{\lambda^2\rho_0^2}{(4\pi)^2}B_0(p^2,m_1^2,0)-\frac{2 \lambda}{(4\pi)^2}\left(1+\frac{1}{6}\frac{p^2}{m_1^2}\right) A(m_1^2)\\
=&  p^2\left(1-\frac{\lambda}{3}C_{\rm UV}\right)\\
&+\frac{8 \lambda m_1^2}{(4\pi)^2}\bigg[\ln\fb{m_1^2}{\mu^2}-\int_0^1 dx\ln\fb{m_1^2-xp^2}{\mu^2}+\frac{1}{24}\frac{p^2}{m_1^2}\left(\ln\fb{m_1^2}{\mu^2}-1\right)\bigg] .
\end{aligned}
\end{align}
For on-shell Goldstone particles, we have
\begin{equation}
\Gamma_{\sigma\sigma}(p^2=0)=0
\end{equation}
in agreement with~\eqref{eq:curved standard goldstone mass on shell}.

As in Appendix~\ref{sec:standard curved}, we calculate the tree-level S matrix element for $\rho\rho\to\sigma\sigma$ in the covariant approach. The contributing diagrams are
\begin{align}
\begin{aligned}
i\mathcal{M}(\rho\rho \to \sigma \sigma) &= \HalfDashedSMatrix\\
&=\HalfDashedFourPoint+\HalfDashedSTree+\HalfDashedTTree+\HalfDashedUTree.
\end{aligned}
\end{align}
As discussed earlier, the ordering in the first diagram does not matter when all particles are on shell. Therefore, we have
\begin{align}
\begin{aligned}
\mathcal{M} (\rho\rho \to \sigma\sigma)&=2 \left(2\lambda-\frac{s}{\rho_0^2}\right)-24 \lambda^2\rho_0^2\frac{1}{s-m_1^2}-16  \lambda^2\rho_0^2\frac{1}{t}-16 \lambda^2\rho_0^2\frac{1}{u}\\
&=-\frac{2 }{\rho_0^2}\left[3\frac{s^2}{s-m_1^2}+2\frac{(m_1^2-t)^2}{t}+2\frac{(m_1^2-u)^2}{u}\right].
\end{aligned}
\end{align}
This result coincides with~\eqref{eq:higgs goldstone standard}.

\section{Example with Linear Potential}
\label{sec:non renorm}

We now consider an example with Lagrangian given by
\begin{equation}
\L=\hf \partial_\mu\rho\partial^\mu\rho+\hf \fb{\rho}{\rho_0}^{2}\partial_\mu\sigma\partial^\mu\sigma-t_\rho \rho-\hf m^2\rho^2.\label{eq:quadratic Lagrangian}
\end{equation}
This has a flat field space -- the kinetic part on its own is just a reparameterisation of two canonical kinetic terms. Additionally, the potential has no interaction terms between the~$\rho$ and~$\sigma$ fields. However, as we shall see, the theory described by~\eqref{eq:quadratic Lagrangian} \emph{is} nonetheless interacting.

The theory has a symmetry-broken vacuum, which we parametrise as
\begin{align}
\langle \rho\rangle =\rho_0 \equiv  -t_\rho  /m^2, 
\qquad
\langle \sigma\rangle =0.
\end{align}
We can then calculate the covariant Feynman rules for this theory using~\eqref{eq:cov feyn rule}.

The propagators are
\begin{align}
\SolidProp =\frac{i}{p^2-m^2}, \qquad \DashedProp =\frac{i}{p^2   },
\end{align}
where a solid line represents the Higgs mode $\rho$ and a dashed line represents the Goldstone mode $\sigma$. 

The three- and four-point interactions are
\begin{align}
\begin{aligned}
\DashedThreePoint&=\frac{i t_\rho  }{\rho_0^2},
\\
\HalfDashedFourPoint&=-2i\frac{t_\rho  }{\rho_0^3},&\DashedFourPoint&=3i\frac{t_\rho}{\rho_0^3},
\end{aligned}
\end{align}
while the five- and six-point interactions are
\begin{align}
\begin{aligned}
\SolidDashedVertex{3}{2}&=6i\frac{t_\rho }{\rho_0^4},&\OneSolidFourDashedVertex&=-9i\frac{t_\rho }{\rho_0^4},
\\
\SolidDashedVertex{4}{2}&=-24i\frac{t_\rho }{\rho_0^5},&\SolidDashedVertex{2}{4}&=36i\frac{t_\rho}{\rho_0^5},&\DashedSixPoint&=-45i\frac{t_\rho }{\rho_0^5}.
\end{aligned}
\end{align}

Notice that there is an infinite series of higher-point vertices, which are proportional to~$t_\rho$. Moreover, these infinite series include interactions that are absent in the standard approach.

To better understand why this theory has an infinite tower of interactions, we switch to a canonical parametrisation and define
\begin{align}
\phi_1 =\rho\cos\fb{\sigma}{\rho_0}, \qquad
\phi_2 =\rho\sin\fb{\sigma}{\rho_0}.
\end{align}
Then,~\eqref{eq:quadratic Lagrangian} takes the form
\begin{equation}
\L=\hf \partial_\mu\phi_1\partial^\mu\phi_1+\hf \partial_\mu\phi_2\partial^\mu\phi_2-\hf m^2(\phi_1^2+\phi_2^2)-t_\rho\sqrt{\phi_1^2+\phi_2^2}.\label{eq:quadratic Lagrangian canonical}
\end{equation}
With this parametrisation, the field space metric becomes manifestly Euclidean and thus ordinary and covariant Feynman rules will be identical. The final term in~\eqref{eq:quadratic Lagrangian canonical} is non-polynomial and thus has an infinite Taylor series expansion. This term therefore leads to an infinite tower of Feynman rules as presented.

\clearpage
\section{Field-Space Riemann Tensor for General Relativity}
\label{sec:configuration riemann}

In Section~\ref{sec:GR}, we have only presented the field-space Ricci tensor and Ricci scalar, but not the full expression for the Riemann tensor $\mathfrak{R}^{(\mu \nu)}{}_{(\alpha \beta)( \rho \sigma )( \gamma \delta)}$ due to its length. In this appendix, we explicitly display $\mathfrak{R}^{(\mu \nu)}{}_{(\alpha \beta)( \rho \sigma )( \gamma \delta)}$. With the aid of the symbolic computer algebra system {\tt Cadabra2}~\cite{Peeters:2007wn, cadabra2}, we find that the field-space Riemann tensor for General Relativity reads
\begin{align}
\begin{aligned}
\mathfrak{R}^{(\mu \nu)}{}_{(\alpha \beta)( \rho \sigma )( \gamma \delta)} =  
&- \frac{1}{32}\delta^{\mu}_{\rho} \delta^{\nu}_{\beta} g_{\sigma \gamma} g_{\alpha \delta} 
- \frac{1}{32}\delta^{\mu}_{\sigma} \delta^{\nu}_{\beta} g_{\rho \gamma} g_{\alpha \delta} 
- \frac{1}{32}\delta^{\mu}_{\beta} \delta^{\nu}_{\sigma} g_{\rho \gamma} g_{\alpha \delta} 
- \frac{1}{32}\delta^{\mu}_{\beta} \delta^{\nu}_{\rho} g_{\sigma \gamma} g_{\alpha \delta} 
\\
&- \frac{1}{32}\delta^{\mu}_{\rho} \delta^{\nu}_{\beta} g_{\sigma \delta} g_{\alpha \gamma} 
- \frac{1}{32}\delta^{\mu}_{\sigma} \delta^{\nu}_{\beta} g_{\rho \delta} g_{\alpha \gamma} 
- \frac{1}{32}\delta^{\mu}_{\beta} \delta^{\nu}_{\sigma} g_{\rho \delta} g_{\alpha \gamma} 
- \frac{1}{32}\delta^{\mu}_{\beta} \delta^{\nu}_{\rho} g_{\sigma \delta} g_{\alpha \gamma} 
\\
&- \frac{1}{32}\delta^{\mu}_{\alpha} \delta^{\nu}_{\rho} g_{\sigma \gamma} g_{\beta \delta} 
- \frac{1}{32}\delta^{\mu}_{\alpha} \delta^{\nu}_{\sigma} g_{\rho \gamma} g_{\beta \delta} 
- \frac{1}{32}\delta^{\mu}_{\rho} \delta^{\nu}_{\alpha} g_{\sigma \gamma} g_{\beta \delta} 
- \frac{1}{32}\delta^{\mu}_{\sigma} \delta^{\nu}_{\alpha} g_{\rho \gamma} g_{\beta \delta} 
\\
&- \frac{1}{32}\delta^{\mu}_{\alpha} \delta^{\nu}_{\rho} g_{\sigma \delta} g_{\beta \gamma} 
- \frac{1}{32}\delta^{\mu}_{\alpha} \delta^{\nu}_{\sigma} g_{\rho \delta} g_{\beta \gamma} 
- \frac{1}{32}\delta^{\mu}_{\rho} \delta^{\nu}_{\alpha} g_{\sigma \delta} g_{\beta \gamma} 
- \frac{1}{32}\delta^{\mu}_{\sigma} \delta^{\nu}_{\alpha} g_{\rho \delta} g_{\beta \gamma}
\\
&+\frac{1}{32}\delta^{\mu}_{\gamma} \delta^{\nu}_{\beta} g_{\rho \delta} g_{\sigma \alpha}
+\frac{1}{32}\delta^{\mu}_{\delta} \delta^{\nu}_{\beta} g_{\rho \gamma} g_{\sigma \alpha}
+\frac{1}{32}\delta^{\mu}_{\beta} \delta^{\nu}_{\delta} g_{\rho \gamma} g_{\sigma \alpha}
+\frac{1}{32}\delta^{\mu}_{\beta} \delta^{\nu}_{\gamma} g_{\rho \delta} g_{\sigma \alpha}
\\
&+\frac{1}{32}\delta^{\mu}_{\gamma} \delta^{\nu}_{\beta} g_{\rho \alpha} g_{\sigma \delta}
+\frac{1}{32}\delta^{\mu}_{\delta} \delta^{\nu}_{\beta} g_{\rho \alpha} g_{\sigma \gamma}
+\frac{1}{32}\delta^{\mu}_{\beta} \delta^{\nu}_{\delta} g_{\rho \alpha} g_{\sigma \gamma}
+\frac{1}{32}\delta^{\mu}_{\beta} \delta^{\nu}_{\gamma} g_{\rho \alpha} g_{\sigma \delta}
\\
&+\frac{1}{32}\delta^{\mu}_{\alpha} \delta^{\nu}_{\gamma} g_{\rho \delta} g_{\sigma \beta}
+\frac{1}{32}\delta^{\mu}_{\alpha} \delta^{\nu}_{\delta} g_{\rho \gamma} g_{\sigma \beta}
+\frac{1}{32}\delta^{\mu}_{\gamma} \delta^{\nu}_{\alpha} g_{\rho \delta} g_{\sigma \beta}
+\frac{1}{32}\delta^{\mu}_{\delta} \delta^{\nu}_{\alpha} g_{\rho \gamma} g_{\sigma \beta}
\\
&+\frac{1}{32}\delta^{\mu}_{\alpha} \delta^{\nu}_{\gamma} g_{\rho \beta} g_{\sigma \delta}
+\frac{1}{32}\delta^{\mu}_{\alpha} \delta^{\nu}_{\delta} g_{\rho \beta} g_{\sigma \gamma}
+\frac{1}{32}\delta^{\mu}_{\gamma} \delta^{\nu}_{\alpha} g_{\rho \beta} g_{\sigma \delta}
+\frac{1}{32}\delta^{\mu}_{\delta} \delta^{\nu}_{\alpha} g_{\rho \beta} g_{\sigma \gamma}
\\
&+\frac{1}{4D} g_{\rho \gamma} g^{\mu \nu} g_{\sigma \beta} g_{\alpha \delta}
+\frac{1}{4D} g_{\rho \delta} g^{\mu \nu} g_{\sigma \beta} g_{\alpha \gamma}
+\frac{1}{4D} g_{\rho \alpha} g^{\mu \nu} g_{\sigma \gamma} g_{\beta \delta}
+\frac{1}{4D} g_{\rho \alpha} g^{\mu \nu} g_{\sigma \delta} g_{\beta \gamma}
\\
&+\frac{1}{4D} g_{\rho \gamma} g^{\mu \nu} g_{\sigma \alpha} g_{\beta \delta}
+\frac{1}{4D} g_{\rho \delta} g^{\mu \nu} g_{\sigma \alpha} g_{\beta \gamma}
+\frac{1}{4D} g_{\rho \beta} g^{\mu \nu} g_{\sigma \delta} g_{\alpha \gamma}
+\frac{1}{4D} g_{\rho \beta} g^{\mu \nu} g_{\sigma \gamma} g_{\alpha \delta} 
\\
&- \frac{1}{4D} g^{\mu \nu} g_{\rho \beta} g_{\sigma \gamma} g_{\alpha \delta} 
- \frac{1}{4D} g^{\mu \nu} g_{\rho \beta} g_{\sigma \delta} g_{\alpha \gamma} 
- \frac{1}{4D} g^{\mu \nu} g_{\rho \gamma} g_{\sigma \alpha} g_{\beta \delta} 
- \frac{1}{4D} g^{\mu \nu} g_{\rho \delta} g_{\sigma \alpha} g_{\beta \gamma} 
\\
&- \frac{1}{4D} g^{\mu \nu} g_{\rho \alpha} g_{\sigma \gamma} g_{\beta \delta} 
- \frac{1}{4D} g^{\mu \nu} g_{\rho \alpha} g_{\sigma \delta} g_{\beta \gamma} 
- \frac{1}{4D} g^{\mu \nu} g_{\rho \delta} g_{\sigma \beta} g_{\alpha \gamma} 
- \frac{1}{4D} g^{\mu \nu} g_{\rho \gamma} g_{\sigma \beta} g_{\alpha \delta}.
\end{aligned}
\end{align}
We note that this tensor vanishes for $D=1$. This is to be expected since the field space of gravity in one dimension cannot be anything other than trivial. 

Note that these results differ from those reported in~\cite{Steinwachs:2017ihd}, where the DeWitt metric was used instead.

\end{appendices}

\end{document}